\documentstyle[epsfig,aps,prb,amstex]{revtex}

\begin{document}

\begin{center}
{\Large {\bf Properties of low-lying states in some high nuclearity Mn, Fe 
and V clusters: Exact Studies of Heisenberg Models}}\\
\vspace{0.5cm}

C. Raghu$^1$, Indranil Rudra$^1$, Diptiman Sen$^2$ and S. Ramasesha$^1$\\
$^1$ Solid State and Structural Chemistry Unit, Indian Institute of Science, 
Bangalore 560012, India\\
$^2$ Centre for Theoretical Sciences, Indian Institute of Science, 
Bangalore 560012, India\\
~\\
~\\
{\bf ABSTRACT}\\
\end{center}

Using an efficient numerical scheme that exploits spatial symmetries and spin 
parity, we have obtained the exact low-lying eigenstates of exchange 
Hamiltonians for the high nuclearity spin clusters, Mn$_{12}$, Fe$_8$ and 
V$_{15}$. The largest calculation involves the Mn$_{12}$ cluster which spans a 
Fock space of a hundred million. Our results show that the earlier estimates 
of the exchange constants need to be revised for the Mn$_{12}$ cluster to 
explain the level ordering of low-lying eigenstates. In the case of the Fe$_8$ 
cluster, correct level ordering can be obtained which is consistent with the 
exchange constants for the already known clusters with butterfly structure. In 
the V$_{15}$ cluster, we obtain an effective Hamiltonian that reproduces 
exactly, the eight low-lying eigenvalues of the full Hamiltonian.

\vspace{.7cm}
\vspace{.5cm}

\section {\bf Introduction}

The synthesis of high nuclearity transition metal complexes has provided a new 
dimension to the field of nanomagnetism. \cite{synth} Many interesting 
phenomena have been observed in these systems. Amongst the most exciting are 
the observation of quantum resonance tunneling and quantum interference 
\cite{qtm} in some of these clusters. For example, in the case of the 
$\rm Mn_{12}$ cluster, the ground state with total spin $S_G=10$ of the 
exchange Hamiltonian, under the influence of a large single ion anisotropy 
gives rise to a manifold of doubly degenerate states with 
nonzero $M_s$ values, with $M_s = \pm 10$ being the lowest energy states. 
The application of a magnetic field splits the degeneracy of the $M_s=\pm 10$ 
states. Varying the magnetic field brings states with $|M_s| \ne 10$ closer in 
energy to the higher of the two states with $|M_s|=10$. The weak spin dipolar 
interactions that exist in the system can connect these nearly degenerate 
states with different $M_s$ values, leading to tunneling between the states.
This is reflected in experiments as jumps in magnetization in the magnetization 
{\it vs} magnetic field plots, whenever the resonance condition is satisfied 
and as plateaus for off-resonance field values. Similar plateaus are also 
observed in the $V_{15}$ cluster \cite{chiorescu}, although the reason for 
the plateaus in this system
is qualitatively different. The quantum interference phenomena observed in
the $\rm Fe_8$ cluster is because the paths connecting the $M_s=+10$
and $M_s=-10$ could interfere in the presence of a magnetic field, leading to 
an oscillation in the tunneling probabilities. \cite{wernsdorfer}

These clusters, at a very basic level are characterized by multidentate ligands
interconnecting the transition metal ions. In the clusters, a given magnetic
ion has exchange interactions of either sign with several of its neighbors. 
Thus, these magnetic clusters often correspond to spin frustrated systems. 
Because of the rather complex exchange pathways which exist in these systems, 
it is difficult to predict a priori even the sign of the exchange constant, 
let alone its magnitude. \cite{sign} Since the site symmetry at the magnetic 
ions is also usually low because of the multidentate ligands in the system, 
the orbital degeneracy would be lifted leading to weaker ferromagnetic exchange 
interactions. Thus one should expect low-spin ground states in these systems.
However, because of the frustrations in the exchange pathways, even weak 
ferromagnetic interactions could lead to higher spin ground states, albeit
with rather low spin excitation gaps. It is indeed interesting that 
the Mn$_{12}$ cluster has ferrimagnetic ordering \cite{qtm} of the spins in its
ground state for this reason. The high-spin ground state observed in Fe$_8$,
is, however, attributable entirely to frustration in the antiferromagnetic 
interactions. \cite{delfs}

In the case of the $\rm Mn_{12}$ cluster, while the ground state spin as well 
as the lowest excitation gap is established experimentally, it is not at all 
clear what the magnitude and sign of the exchange interactions in the cluster 
are. In an earlier study \cite{gat1}, in order to simplify the calculations,
each strongly coupled $\rm Mn^{III}-Mn^{IV}$ pair was replaced by a composite 
spin-$1/2$ object. The resulting model was studied for three different sets 
of exchange constants. It was observed that the ordering of the energy levels 
is very sensitive to the variations in the exchange constants. 

In the case of the $\rm Fe_8$ cluster, while model exact calculations 
\cite{delfs} were
possible because of the smaller dimensions of the Hilbert spaces, the exchange
parameters used were very different from those that have been determined
recently. \cite{fe8jval} Considering the sensitivity of the ordering of the 
energy levels to values of the exchange constants, it is desirable to redo 
the calculations using revised estimates of the exchange constants.

The simplest cluster that can be studied exactly is the ${\rm V_{15}}$ cluster.
In this cluster it is found that eight low-lyings states are well separated 
from the rest of the spectrum. \cite{v15} Most of the low-temperature 
properties are determined by these eight low-lying states. To undertake 
serious study of the magnetization behavior of the cluster under the influence 
of an applied magnetic field, it is necessary to construct an accurate model 
Hamiltonian for these states. 

In this paper, we have studied the low-lying states of $\rm Mn_{12}, ~ Fe_8~ 
and ~ V_{15}$ clusters using exact diagonalization of the corresponding
exchange Hamiltonians. We show that in the case of $\rm Mn_{12}$, earlier
estimates of the exchanges constants fail to provide the ground and excited
state spin quantum numbers in agreement with experiments. We have also 
estimated the most likely exchange constants that give good agreement with 
experiments. In the case of the $\rm Fe_8$ cluster, we have studied the 
properties of the cluster for more recent estimates of the exchange constant.
In the case of $\rm V_{15}$ we have obtained an effective Hamiltonian
for the low-lying states that reproduces the energy level ordering of the
eight low-lying states exactly. Such a Hamiltonian would be important in the
context of hysteresis studies of the system at low temperatures.

In the next section, we outline the numerical method for obtaining the
low-lying states of exchange coupled spin systems which span large Hilbert
spaces. In Sec. 3, we discuss the results obtained from this technique for the 
three magnetic clusters mentioned above, and we summarize our results in Sec. 
4.

\section{\bf Model Hamiltonian and Computation Details}

The model Hamiltonian employed in these studies is the isotropic exchange
Hamiltonian involving exchange interactions between nearest neighbors, 
\begin{equation}
{\hat H} ~=~ \underset{<ij>}{\Sigma} ~J_{ij} ~{\hat s}_i \cdot {\hat s}_j ~,
\end{equation}
where the exchange interaction $J_{ij}$ takes the values dictated by 
experimental studies of structure and magnetic properties.
The total dimensionality of the Fock space of the cluster is given by
\begin{equation}
D_F ~=~ \overset{n}{\underset{i=1}{\Pi}} ~(2S_i+1)~,
\end{equation}
where $n$ is the total number of spins in the cluster and $S_i$ is the spin 
on each ion. In the case of the $\rm Mn_{12}$ cluster consisting of eight
spin-$2$ ions and four spin-$3/2$ ions, the Fock space dimensionality is a 
hundred million. Specializing to a given total $M_S$ leads to Hilbert space 
dimensionalities which are lower than the Fock space dimensionality. In the
case of the $\rm Mn_{12}$ cluster the $M_S=0$ space has a dimensionality
of over eight million (8,581,300). The major challenge in exact computation
of the eigenvalues, and properties of these spin clusters lies in handling
such large basis and the associated matrices. While the dimensions look 
overwhelming, the matrices that represent the operators in these spaces are
rather sparse. Usually, the number of nonzero elements in a row is of the
order of the number of exchange constants in the Hamiltonian. This sparseness
of the matrices allows one to handle fairly large systems. However, in the
case of spin problems, generating the basis states and using the symmetries
of the problem is nontrivial.

The isotropic exchange Hamiltonians conserve the total spin, $S$, besides the 
$z$-component of the total spin, $M_S$. Besides these symmetries, the geometry
of the cluster also leads to spatial symmetries which can often be exploited.
The simplest way of generating basis functions which conserve total spin is 
the VB method that employs the Rumer-Pauling rule. \cite{vb} It is quite easy 
to generalize the Rumer-Pauling rules to a cluster consisting of objects 
with different spins to obtain states with desired total spin, $S$. However,
setting up the Hamiltonian matrix in such a basis can be computer intensive 
since the exchange operators operating on a "legal" VB diagram (diagram that 
obeys Rumer-Pauling rules) can lead to "illegal" VB diagrams, and resolving
these "illegal" VB diagrams into "legal" diagrams would present the major
bottle-neck. Indeed, the same difficulty is encountered when spatial symmetry
operators operate on a VB function. Thus, the extended VB methods are not
favored whenever one wishes to apply it to a motley collection of spins or 
when one wishes to exploit some general spatial symmetries that may exist in 
the cluster.

Usually, in frustrated spin systems, it is important to partition the spaces 
into different total spin spaces because of the usually small energy gaps 
between total spin states which differ in $S$ by unity. To avoid the 
difficulties involved in working with total spin eigenfunctions, we exploit 
parity symmetry in the systems. The parity operation involves changing the
$z$-component of all the spins in the cluster from $M_{S_i}$ to $-M_{S_i}$. 
There is an associated phase factor with this operation given by $(-1)^{
S_{tot} + \sum_i S_i}$. The isotropic exchange operator remains invariant 
under this operation. If this symmetry is employed in the $M_S=0$ subspace, 
the subspace is divided into "even" and "odd" parity spaces depending upon 
the sign of the character under the irreducible representation of the parity 
group. The space which corresponds to even (odd) total spin we call the even 
(odd) parity space. Thus, employing parity allows partial spin symmetry 
adaptation which separates successive total spin spaces, without introducing
the complications encountered in the VB basis. However, the VB method can lead 
to complete factorization of the spin space leading to smaller complete
subspaces.

In the $\rm Mn_{12}$ cluster, besides spin symmetries, there also exists 
spatial symmetries. The topology of the exchange interaction leads to a $C_4$
point group symmetry. At first sight, this point group appears to present
difficulties because the characters in the irreducible representation are
in some cases complex. This could lead to complex basis functions. This, 
however, can be avoided by recognizing that in the $C_n$ group, states with
wave vectors $k$ and $-k$ are degenerate in the absence of an external magnetic
field. We can therefore construct a linear combination of the $k$ and $-k$
states which is real. The symmetry representations in the $C_4$ group 
would then correspond to the labels $A$, $B$ and $E$, with the characters in 
the $E$ representation given by ${\rm 2cos} (rk)$ under the symmetry operation
$C^r_4$, with $k=\pi / 2$. The parity operation commutes with the spatial
symmetry operations, and the full point group of the system would then 
correspond to the direct product of the two groups. Since both parity and 
spatial symmetries can be easily incorporated in a constant $M_S$ basis, we
do not encounter the difficulties endemic to the VB theory.

The generation of the complete basis in a given Hilbert space requires a simple
representation of a state on the computer. This is achieved by associating 
with every state a unique integer. In this integer, we associate $n_i$ bits
with spin $s_i$, such that $n_i$ is the smallest integer for which $2^{n_i}
\ge 2s_i+1$. In the integer that represents the state of the cluster, we
ensure that these $n_i$ bits do not take values which lead to the $n_i$-bit 
integer value exceeding $2s_i+1$. For each of the allowed bit states of
the $n_i$-bit integer, we associate an $M_{S_i}$ value between $-s_i 
{\rm ~and}~ s_i$. For a spin cluster of $n$ spins, we scan all integers of 
bit length $N=\overset{n}{\underset{i=1}{\Sigma}}n_i$ and verify if it 
represents a basis state with the desired $M_S$ value. In Fig. 1, we show a 
few basis functions with specified $M_s$ value for some typical clusters along 
with their bit representations and the corresponding integers. Generation of 
the basis states is usually a very fast step, computationally. Generating the 
basis as an ordered sequence of integers that represent them also allows for 
a rapid generation of the Hamiltonian matrix elements as will be seen later.

Symmetrization of the basis by incorporating parity and spatial symmetries 
involves operating on the constant $M_S$ basis by the symmetry operators. 
Since spatial symmetry operators exchange the positions of equivalent spins,
every spatial symmetry operator operating on a basis function generates 
another basis function. Every symmetry operator can be represented by a 
correspondence vector whose $i^{\rm th}$ entry gives the state that results 
from operating on the $i^{\rm th}$ state by the chosen operator. This is also 
true for the parity operator, in the $M_S~=~0$ subspace. The symmetry 
combinations can now be obtained by operating on each state by the group 
theoretic projection operator,
\begin{equation}
{\hat P}_{\Gamma_i}~=~ {{1} \over {h}}~ \underset{R}{\Sigma}~
\chi_{\Gamma_i}(R) ~{\hat R}
\end{equation}
on each of the basis states. Here $\Gamma_i$ is the $i^{\rm th}$ irreducible 
representation, ${\hat R}$ is the symmetry operation of the group, $h$ is the
order of the symmetry group, and
$\chi_{\Gamma_i}(R)$ is the character under ${\hat R}$ in the irreducible
representation $\Gamma_i$. The resulting symmetrized basis is overcomplete. 
The linear dependencies can be eliminated by a Gram-Schmidt orthonormalization
procedure. However, in most cases, ensuring that a given basis function does 
not appear more than once in a symmetrized basis is sufficient to guarantee
linear independence and weed out the linearly dependent states. A good check
on the procedure is to ensure that the dimensionality of the symmetrized
space agrees with that calculated from the traces of the reducible 
representation obtained from the matrices corresponding to the symmetry
operators. Besides, the sum of the dimensionalities of the symmetrized spaces 
should correspond to the dimensionality of the unsymmetrized Hilbert space.

The generation of the Hamiltonian matrix is rather straightforward and 
involves operation of the Hamiltonian operator on the symmetry 
adapted basis. This results in the matrix ${\mathbf SH}$, where ${\mathbf S}$ 
is the symmetrization matrix representing the operator ${\hat P}_{\Gamma_i}$ 
and ${\mathbf H}$ is the matrix whose elements $h_{ij}$ are defined by 
\begin{equation}
{\hat H} |i> = \underset{j}{\Sigma}~ h_{ij} |j>.
\end{equation}
The states $i$ correspond to the unsymmetrized basis functions. The 
Hamiltonian matrix in the symmetrized basis is obtained by right multiplying
the matrix ${\mathbf SH}$ by ${\mathbf S^\dagger}$. The symmetric
Hamiltonian matrix is stored in the sparse matrix form and the matrix 
eigenvalue problem is solved using the Davidson algorithm. 

Computation of the properties is easily done by transforming the eigenstate 
in the symmetrized basis into that in the unsymmetrized basis. Since the
operation by any combination of spin operators on the unsymmetrized basis 
can be carried out, all relevant static properties in different eigenstates
can be obtained quite simply.

\section{\bf Results and Discussion}

We have solved the exchange Hamiltonian exactly for the ${\rm Mn_{12},~Fe_8~
and ~V_{15}}$ clusters using the above mentioned method. We have obtained the
eigenvalues and various properties of the eigenstates such as spin densities
and spin-spin correlation functions for these clusters. In what follows, we
will discuss these in detail.

\subsection{\bf Mn$_{12}$Ac Cluster}

In Fig. 2 we show the geometry and the exchange parameters for this
cluster. The crystal structure suggests that the exchange constant $J_1$ is
largest and antiferromagnetic in nature. \cite{mn12ac1} Based on magnetic 
measurements, it has been suggested that $J_1$ has a magnitude of 215K. The 
other magnitude and sign of the other exchange constants are based on 
comparisons with manganese systems in smaller clusters. \cite{mn12ac1} 
It has been suggested that the exchange 
constant $J_2$ and $J_3$ are antiferromagnetic and have a magnitude of about
85K. However, for the exchange constant $J_4$, there is no concrete estimate,
either of the sign or of the magnitude. In an earlier study, the Mn$^{III}$ - 
Mn$^{IV}$ pair with the strongest antiferromagnetic exchange constant was 
replaced by a composite spin-$1/2$ object \cite{gat1}, and the exchange 
Hamiltonian of the cluster was solved for three different sets of parameters.
It was found that the ordering of the energy levels were very sensitive to
the relative strengths of the exchange constants. In these studies, $J_4$ was 
set to zero and the low-lying excited states were computed. Besides, only 
states with spin $S$ up to $10$ could be obtained because of the 
replacement of the higher spin ions by composite spin-$1/2$ objects.

In our calculation, we have dealt with all the magnetic ions in the cluster 
and using symmetry, we have factored the $M_S=0$ Hilbert space into the 
six symmetry subspaces. The dimensionalities of the different subspaces is 
given in Table I. We have obtained low-lying eigenstates in each of these 
sectors and determined the total spin of the state by explicitly computing 
the expectation value of the ${\hat S}^2$ operator in the state. 

Our results for the low-lying states are shown in Table II. We note that none 
of the three sets of parameters studied using an effective Hamiltonian, gives 
the correct ground and excited states, when an exact calculation is performed.
It appears that setting the exchange constant $J_4$ to zero, cannot yield
an $S=10$ ground state (Table II, cases A, B and C). When $J_3$ is equal to
or slightly larger than $J_2$ (cases A and B, Table II), we find a singlet 
ground state, unlike the result of the effective Hamiltonian in which the 
ground state has $S=8$ and $S=0$ respectively. The ground state has spin $S=6$, 
when $J_3$ is slightly smaller than $J_2$ (case C, Table II). In all these 
cases, the first few low-lying states are found to lie within $20$K of the 
ground state.

When we use the parameters suggested by Chudnovsky \cite{chudnov} (case D, 
Table II), we obtain an $S=10$ ground state separated from an $S=9$ first 
excited state by 223K. This is followed by another $S=9$ excited state at 
421K. Only when the exchange constant $J_4$ is sufficiently strongly 
ferromagnetic (case E, Table II), do we find an $S=10$ ground state with an 
$S=9$ excited state separated from it by a gap of 35K, which is close to the 
value inferred indirectly from experimental results. \cite{egap} The second 
higher excited state has $S=8$, and is separated from the ground state by 62K. 

We have explored the parameter space a little further by varying $J_3$ and 
$J_4$, to see the effect of these exchange constants on the ordering of the 
energy levels. We find that for $|J_3|=|J_4|$ and $J_3$ antiferromagnetic 
but $J_4$ ferromagnetic, the ground state is always $S=10$ (Table III, cases 
C, D and E); the first and second excited states are $S=9$ and $S=8$, 
respectively. The lowest excitation gap decreases slowly with increasing 
magnitude of the exchange constants.

We find that the spin of the ground state is very sensitive to $J_4$, for 
fixed value of $J_3$. In the case where $J_4$ weakly ferromagnetic (Table III, 
case B), we obtain an $S=0$ ground state, $J_4$ weakly antiferromagnetic we 
obtain an $S=4$ ground state (Table III, case A). This shows that frustrations 
play a dominant role. If $J_3$ is also made ferromagnetic, the role of 
frustration is considerably reduced. 

In Figs. 3(a) and 3(b), we show the spin density \cite{mn12spnden} for the 
Mn$_{12}$ cluster in the ground state for the $S=10,M_S=10$ state. While the 
manganese ions connected by the strong antiferromagnetic exchange show 
opposite spin densities, it is worth noting that the total spin density on 
these two ions is $0.691$, well away from the value of $0.5$ expected, if 
these ions were indeed to form a spin-$1/2$ object. We also note that the spin 
density at the manganese ion in the middle of the crown is much larger than 
that at the corners. The spin density in the excited state $S=9, M_S=9$, also 
has a similar distribution, although in this state, the symmetry of the spin 
Hamiltonian is apparently broken [Fig. 3(b)]. The corner ions in the crown no 
longer have the same spin densities; one pair of opposite corner ions have 
larger spin densities than the other corner pair. However, since 
this state is doubly degenerate, there is another state in which the spin 
densities are related to the spin densities of this state by a $90^0$ 
rotation. In any experiment involving this state, only an arbitrary linear 
combination of the two spin densities will be observed. Note also that the 
large differences in the spin densities for the closely lying excited states 
is an indication of the large degree of spin frustration in the system.

The small energy gap (35K) between the $S=10$ ground state and 
the $S=9$ lowest excited state seems to indicate that, if the $g$ factors
of the Mn ions in the core and crown are different then, an applied magnetic
field should mix these spin states. Such a mixing would then be reflected 
in the quantum resonance tunneling experiments. However, it appears that the
experiments are well described by the $S=10$ state alone. This is what
we should expect from the symmetry of the two low-lying states. We note
that the ground state has $A$ symmetry while the lowest excited state has
a $E$ symmetry. These two states cannot be mixed by any perturbation that
retains the $C_4$ symmetry of the cluster.

\subsection{\bf Fe$_8$ Cluster}

The Fe$_8$ cluster is shown in Fig. 4. Each of the Fe ions has a spin 
of 5/2 and the ground state of the system has a total spin $S=10$, with $S=9$
excited state separated from it by about 20K. All the exchange interactions
in this system are expected to be antiferromagnetic. While the structure of the 
complex dictates that the exchange interaction $J_2$ along the back of the
butterfly should be considerably smaller than the interaction $J_1$ across the 
wing \cite{fe8butterfly}, in earlier studies it was reported that such a 
choice of interaction parameters would not provide a $S=10$ ground state. 
\cite{delfs} 

We have carried out exact calculations of the eigenstates of the Fe$_8$ cluster 
using four sets of parameters; the last set of parameters (Case D) are taken
from Ref. [17]. In cases A, C and D, $J_2$ is very much smaller 
than $J_1$. We find that in all the four cases, the ground state has a 
spin $S=10$ and the lowest excited state has spin $S=9$. One of the main 
differences we find amongst the four sets of parameters is in the energy gap 
to the lowest excited state (Table IV). For the set of parameters used in the 
earlier study, this gap is the lowest at $3.4$K (case B). For the parameter 
sets A, C and D \cite{fe8jval}, this gap is respectively $13.1$K, $39.6$K
and $42.4$K. While in cases A, C and D, the second excited state has spin-$9$, 
in case B, this state has spin-$8$. 

The spin densities in all the four cases for both the ground and the excited 
state are shown in Figs. 5(a) to 5(h). The spin densities in all cases are 
positive at the corners. In cases A and B, the spin density is negative on the 
Fe ions on the backbone, and is positive on the remaining two Fe sites. 
\cite{fe8spnden,barra} However, in cases C and D, the negative 
and positive spin 
density sites for the Fe ions in the middle of the edges are interchanged. This 
is perhaps due to the fact that in cases A and B, the exchange constant $J_3$ 
is less than $J_4$, while in cases C and D, this is reversed. Thus, a spin 
density measurement can provide relative strengths of these two exchange 
constants. In all the cases, the difference between the spin densities in the 
ground and excited states is that the decrease in the spin density in the 
excited state is mainly confined to the corner Fe sites. Note that the spin
densities in cases C and D are almost the same, although the excitation gaps
are significantly different and in proportion to the differences in the
exchange constants.

We should note here that the spin densities presented by us are expectation 
values of the site operators $S_i^z$ in the $S=10, M_S =10$ and $S=9, M_S =9$ 
states. However, experimental values are obtained not only at the Fe sites but 
also at the spin polarized neighboring ligand atom sites. We therefore use the 
experimental results only as a guideline for locating the negative spin 
density Fe sites which are sensitive to the set of exchange constants used in 
the calculation. 

The Fe$_8$ cluster is quite different from the Mn$_{12}$ cluster in the
following sense. In the Fe$_8$ cluster, we have excited states of spin $S=8$
and $S=9$ which have the same symmetry as that of the $S=10$ ground state.
Furthermore, the total splitting of the ground state due to the anisotropic 
terms arising in the system due to spin-dipolar interactions is larger than 
the energy gaps with the $S=8$ and $S=9$ states of the same spatial symmetry 
as the ground state. Thus, if the $g$ factors of the Fe ions on the backbone 
of the butterfly are different from those on the wings, then an applied
magnetic field could lead to mixing between the different spin states. We
expect this to provide an additional mechanism for quantum resonance tunneling 
in the Fe$_8$ cluster. 

\subsection{\bf V$_{15}$ Cluster}

The simplest cluster to study is the V$_{15}$ cluster, since each of the ions
has a spin of half. The interesting aspect of the V$_{15}$ cluster is that the 
three spins sandwiched between the hexagons [Fig. 6] have no direct spin-spin 
interactions. All the interactions shown in Fig. 6 are antiferromagnetic and 
the spin system is frustrated. The eigenstates of this system consist of eight 
states corresponding to the triangle spins (i.e., three spin-$1/2$ sites) which 
are split off from the rest of the spectrum. A combination of three two-spin 
interactions which retains the $C_3$ symmetry of the molecule is sufficient to 
account for such a spectrum. We find that the effective Hamiltonian is given by
\begin{equation}
H_{sp-sp}~ =~ \epsilon ~I ~+ ~\alpha ~(S_1 \cdot S_2 + S_2 \cdot S_3 + S_3
\cdot S_1) ~,
\label{eff}
\end{equation}
where $\epsilon = -4.78187$ and $\alpha = 0.02015$ in 
units of the exchange $J_1$. This Hamiltonian reproduces the eight low-lying 
eigenstates of the full exchange Hamiltonian to numerical accuracy. 

The spin density distribution in one of the $S=1/2, M_S=1/2$ ground states
as well as in the $S=3/2, M_S=3/2$ excited state is shown in Figs. 7(a) 
and 7(b). We find that the spin densities on the hexagons are negligible. 
The total spin density in the triangle for the $S=1/2$ state is nearly equal 
in value to that a free electron spin. The $S=3/2, M_S=3/2$ state also has 
almost equal spin densities at all three sites of the triangle, nearly equal 
to that of free spins. These observations suggest that describing the 
low-energy spectrum of this system by the triangle spins is quite appropriate. 

\section{\bf Summary}

To conclude, using a bit representation of the spin states of a spin cluster
combined with exploitation of spatial symmetry and spin parity, we are able
to obtain model exact solutions for exchange Hamiltonians whose Fock space
spans up to a hundred million states. Our studies on the Mn$_{12}$Ac cluster
cluster shows that the earlier effective Hamiltonian studies wrongly estimated
the exchange constants. The new exchange constants give the correct spin for 
the ground state as well as the correct ordering of the low-lying excited 
states. The spin densities in the cluster also support the fact that the 
effective Hamiltonian in earlier studies does not accurately represent the 
cluster. The studies on Fe$_8$ cluster shows that the correct ground state as 
well as the first excited state can be obtained by using a set of exchange 
constants consistent with the butterfly structure known in related systems, 
unlike what was concluded in an earlier study. The interesting feature of the 
Fe$_8$ spin densities is that the sites which have negative spin densities 
depend on the relative strengths of the exchange constants on the sides of the 
butterfly. In the case of the V$_{15}$ cluster, we find that the eight 
low-lying energy levels can be fitted to an effective Hamiltonian describing 
three spin-$1/2$ sites.

\newpage

\begin{center}
{\bf Acknowledgments}\\
\end{center}

We thank the Council of Scientific and Industrial Research, India for 
financial support through grant No. 01(1595)/99/EMR-II.

\newpage

\noindent
Table I: Dimensions of different symmetry subspaces in $\rm M_s$=0 sector for 
$\rm Mn_{12}Ac$.

\begin{center}
\begin{tabular}{|c|c|} \hline
{\bf Symmetry} & {\bf Dimension} \\ \hline
$^e$A & 1074087 \\ 
$^o$A & 1071537 \\ 
$^e$B & 1074037 \\ 
$^o$B & 1071587 \\ 
$^e$E & 2142526 \\ 
$^o$E & 2147526 \\ \hline
\end{tabular}
\end{center}

\vspace{2cm}

\noindent
Table II: Low-Lying states of $\rm Mn_{12}Ac$, relative to the ground state
for the parameters in question. Entries in parenthesis in
cases A, B and C correspond to the effective Hamiltonian results of Sessoli
{\it et al}. \cite{gat1} Case D corresponds to the parameters suggested by
Chudnovsky. \cite{chudnov} The parameters corresponding to different cases are:
case (A) $J_1$=225K, $J_2$=90K, $J_3$=90K, $J_4$=0K; case (B) $J_1$=225K, 
$J_2$=90K, $J_3$=93.8K, $J_4$=0K; case (C) $J_1$=225K, $J_2$=90K, $J_3$=86.2K, 
$J_4$=0K; case (D) $J_1$=215K, $J_2$=85K, $J_3$=-85K, $J_4$=-45K; case (E) 
$J_1$=215K, $J_2$=85K, $J_3$=85K, $J_4$=-64.5K. All the energies are in K.

\begin{center}
\begin{tabular}{|c|c|c|c|c|}
\hline 
\hspace{0cm} {\bf Case A} \hspace{0cm} & \hspace{0cm} {\bf Case B}
\hspace{0cm} & \hspace{0cm} {\bf Case C} \hspace{0cm} &
\hspace{0cm} {\bf Case D} \hspace{0cm} & \hspace{0cm} {\bf Case E} 
\hspace{0cm} \\ \hline 
State ~~~~ S ~~~~ E(K) & State ~~~~S ~~~~ E(K) & State ~~~~S ~~~~ E(K) &
State ~~~~S ~~~~ E(K) & State ~~~~S ~~~~E(K) \\ \hline 
$^e{B}$~~~~0~~~~ 0.0 & $^e{B}$~~~~0~~~~ 0.0 & $^e{B}$~~~~6~~~~ 0.0 
& $^e{A}$~~~~10~~~~ 0.0 & $^e{A}$~~~~10~~~~ 0.0 \\ 

(8) &(0) & (10) & & \\ \hline

$^o{E}$~~~~ 1~~~~10.8 & $^o{E}$ ~~~~ 1~~~~16.2 & $^o{E}$ ~~~~ 1~~~~15.5 
& $^o{E}$ ~~~~ 9~~~~223 & $^o{E}$ ~~~~ 9~~~~35.1 \\

~~~~~~(9)~~~~(6.4) &~~~~~~(8)~~~~(1.4) &~~~~~~(8)~~~~(2.7) & & \\ 

~~~~~~(10)~~~~(6.4) & & & & \\ \hline

$^o{B}$ ~~~~ 1~~~~19.8 & $^o{B}$ ~~~~ 1~~~~20.0 & $^o{B}$ ~~~~1~~~~19.6 
& $^o{B}$ ~~~~9~~~~421.2 & $^e{B}$ ~~~~8~~~~62.1 \\

~~~~~~(0)~~~~(6.8) & &~~~~~~(9)~~~~(5.0) & & \\ \hline

$^e{A}$ ~~~~ 2~~~~24.7 & $^e{A}$ ~~~~ 2~~~~30.5 & $^e{A}$ ~~~~2~~~~23.8 
& $^o{B}$ ~~~~9~~~~425.1 & $^o{E}$ ~~~~7~~~~82.4 \\ \hline

$^o{E}$ ~~~~3~~~~39.0 & $^e{B}$ ~~~~4~~~~58.4 & $^o{E}$ ~~~~1~~~~28.8 
& $^e{B}$ ~~~~ 8~~~~439.5 & $^e{A}$ ~~~~6~~~~99.7 \\ \hline

$^e{E}$ ~~~~2~~~~49.9 & $^e{E}$ ~~~~2~~~~60.9 & $^e{B}$ ~~~~6~~~~53.6 
& $^e{B}$ ~~~~8~~~~443.7 & $^e{B}$ ~~~~0~~~~102.0 \\ \hline

$^e{B}$ ~~~~4~~~~57.1 & $^o{A}$ ~~~~3~~~~64.3 & $^e{B}$ ~~~~6~~~~54.4 
& $^e{B}$ ~~~~8~~~~458.1 & $^e{A}$ ~~~~2~~~~121.0 \\ \hline

$^e{B}$ ~~~~8~~~~57.8 & $^e{E}$ ~~~~2~~~~80.0 & $^e{B}$ ~~~~8~~~~57.2 
& $^o{A}$ ~~~~11~~~~573.4 & $^0{B}$ ~~~~1~~~~133.3 \\ \hline

$^e{B}$ ~~~~2~~~~57.8 & $^o{A}$ ~~~~3~~~~88.1 & $^e{E}$ ~~~~2~~~~63.0 
& $^o{E}$ ~~~~9~~~~583.8 & $^e{E}$ ~~~~2~~~~177.1 \\ \hline

$^o{B}$ ~~~~3~~~~78.4 & $^e{A}$ ~~~~6~~~~88.3 & $^o{A}$ ~~~~3~~~~77.0 
& $^e{E}$ ~~~~8~~~~632.8 & $^o{A}$ ~~~~3~~~~211.3 \\ \hline

$^o{B}$ ~~~~3~~~~86.8 & $^o{B}$ ~~~~3~~~~112.8 & $^o{B}$ ~~~~3~~~~85.3 
& $^o{A}$ ~~~~9~~~~ 640.5 & $^o{A}$ ~~~~3~~~~220.8 \\ \hline

$^e{A}$ ~~~~6~~~~105.7 & $^o{B}$ ~~~~5~~~~ 114.6 & $^e{E}$ ~~~~2~~~~86.1 
& $^e{E}$ ~~~~8~~~~658.3 & $^e{E}$ ~~~~4~~~~249.9 \\ \hline

$^o{B}$ ~~~~3~~~~113.4 & $^o{B}$ ~~~~5~~~~158.4 & $^e{A}$ ~~~~6~~~~97.1 
& $^e{A}$ ~~~~8~~~~767.1 & $^0{B}$ ~~~~5~~~~278.5 \\ \hline

$^e{E}$ ~~~~4~~~~117.3 & $^o{A}$ ~~~~1~~~~165.2 & $^e{A}$ ~~~~6~~~~98.2 
& $^e{B}$ ~~~~8~~~~807.6 & $^o{A}$ ~~~~7~~~~332.1 \\ \hline

$^o{B}$ ~~~~5~~~~154.2 & $^o{A}$~~~~ 1~~~~181.6 & $^o{B}$~~~~ 3~~~~112.2 
& $^e{A}$~~~~ 8~~~~815.8 & $^o{A}$~~~~ 7~~~~340.8 \\ \hline
\end{tabular} 
\end{center}

\pagebreak
\clearpage

\noindent
Table III: Low Lying states of $\rm Mn_{12}Ac$. The parameters corresponding to 
different cases are: case (A) $J_1$=215K, $J_2$=85K, $J_3$=85K, $J_4$=45K; 
case (B) $J_1$=215K, $J_2$=85K, $J_3$=85K, $J_4$=-45K; case (C) $J_1$=215K, 
$J_2$=85K, $J_3$=64.5 K, $J_4$=-64.5K; case (D) $J_1$=215K, $J_2$=85K, 
$J_3$=85K, $J_4$=-85K; case (E) $J_1$=215K, $J_2$=85K, $J_3$=45K, $J_4$=-45K. 
All the energies are in K.

\begin{center}
\begin{tabular}{|c|c|c|c|c|}
\hline 
\hspace{0cm} {\bf Case A} \hspace{0cm} & \hspace{0cm} {\bf{Case B}}
\hspace{0cm} & \hspace{0cm} {\bf{Case C}} \hspace{0cm} &
\hspace{0cm} {\bf{Case D}} \hspace{0cm} & \hspace{0cm} {\bf{Case E}} 
\hspace{0cm} \\ \hline 
State~~~S~~E(K) & State~~~S~~E(K) & State~~~S~~E(K) &
State~~~S~~E(K) & State~~~S~~E(K) \\ \hline 
$^e{B}$~~~~4~~ 0.0 & $^e{B}$~~~~0~~ 0.0 & $^e{A}$~~~~10~~ 0.0 
& $^e{A}$~~~~10~~ 0.0 & $^e{A}$~~~~10~~ 0.0 \\ \hline

$^e{A}$~~~~4~~9.1 & $^o{E}$~~~~1~~12.3 & $^o{E}$~~~~9~~73.7 
& $^o{E}$~~~~9~~67.7& $^o{E}$~~~~9~~80.1 \\ \hline

$^o{E}$~~~~3~~9.4 & $^e{A}$~~~~2~~22.9 & $^e{B}$~~~~8~~135.1 
& $^e{B}$~~~~8~~121.2 & $^e{B}$~~~~8~~149.8 \\ \hline

$^e{B}$~~~~4~~18.2 & $^0{B}$~~~~1~~27.6 & $^o{E}$~~~~7~~186.1
& $^o{E}$~~~~7~~~165.2 & $^e{A}$~~~~8~~~191.0 \\ \hline

$^e{A}$~~~~2~~32.4 & $^o{E}$~~~~3~~28.9 & $^e{A}$~~~~8~~~196.0 
& $^e{A}$~~~~6~~201.2 & $^o{E}$~~~~7~~~210.0 \\ \hline

$^o{B}$~~~~5~~49.4 & $^e{B}$~~~~4~~34.1 & $^e{A}$~~~~6~~227.8
& $^e{A}$~~~~8~~206.5 & $^e{A}$~~~~6~~260.0 \\ \hline

$^e{A}$~~~~6~~50.0 & $^e{A}$~~~~10~~36.5 & $^e{B}$~~~~4~~283.5
& $^e{B}$~~~~4~~247.7 & $^e{B}$~~~~4~~329.8 \\ \hline

$^e{E}$~~~~4~~55.4 & $^e{B}$~~~~8~~37.8 & $^o{B}$~~~~1~~323.0
& $^o{b}$~~~~1~~282.5 & $^0{B}$~~~~9~~346.8 \\ \hline

$^o{A}$~~~~3~~68.2 & $^e{E}$~~~~2~~67.2 & $^e{E}$~~~~2~~364.0 
& $^e{E}$~~~~2~~330.2 & $^o{B}$~~~~9~~370.7 \\ \hline

$^o{A}$~~~~3~~70.2 & $^o{A}$~~~~3~~100.1& $^o{A}$~~~~3~~391.8
& $^o{A}$~~~~3~~365.2 & $^o{B}$~~~~1~~515.8 \\ \hline

$^o{B}$~~~~3~~71.4 & $^o{A}$~~~~3~~119.5 & $^o{A}$~~~~3~~401.6
& $^o{A}$~~~~3~~ 375.0 & $^e{E}$~~~~8~~400.3 \\ \hline

$^o{A}$~~~~3~~76.6 & $^e{A}$~~~~4~~ 140.0 & $^e{E}$~~~~4~~420.6
& $^e{E}$~~~~4~~401.9 & $^e{E}$~~~~2~~413.8 \\ \hline

$^e{B}$~~~~2~~255.2 & $^o{B}$~~~~3~~161.8 & $^o{B}$~~~~9~~426.3
& $^o{B}$~~~~11~~421.0 & $^o{A}$~~~~5~~424.2 \\ \hline

$^e{B}$~~~~2~~257.2 & $^o{B}$~~~~5~~172.8 & $^o{B}$~~~~5~~434.9
& $^o{B}$~~~~5~~425.5 & $^o{A}$~~~~3~~432.5 \\ \hline

\end{tabular} 
\end{center}

\vspace{2cm}

\noindent
Table IV: Energies (in units of K) of a few low-lying states in 
${\rm Fe_8}$. The exchange constants corresponding to the various cases
are: case (A) J$_1$ = 150K, J$_2$ = 25K, J$_3$ = 30K, J$_4$ = 50K; 
case (B) J$_1$ = 180K, J$_2$ = 153K, J$_3$ = 22.5K, J$_4$ = 52.5K;
case (C) J$_1$ = 195K, J$_2$ = 30K, J$_3$ = 52.5K, J$_4$ = 22.5K; 
case (D) J$_1$ = 201K, J$_2$ = 36.2K, J$_3$ = 58.3K, J$_4$ = 26.1K.

\begin{center}
\begin{tabular}{|c|c|c|c|c} 
\hline
\hspace{0cm} {\bf Case A} \hspace{0cm} & \hspace{0cm} {\bf Case B} 
\hspace{0cm} & \hspace{0cm} {\bf Case C} \hspace{0cm} & \hspace{0cm} 
{\bf Case D} \hspace{0cm} \\ \hline
~~~State ~~~~~~S~~~~~~ E(K) & ~~~State ~~~~~~ S~~~~~~ E(K) & ~~~
State ~~~~~~ S~~~~~~ E(K) & ~~~ State~~~~~~ S~~~~~~ E(K) \\ \hline

$^e{A}$~~~~~~~~~ 10 ~~~~~~~~~ 0.0 & $^e{A}$~~~~~~~~~ 10~~~~~~~~~ 0.0
& $^e{A}$~~~~~~~~~ 10~~~~~~~~~ 0.0 & $^e{A}$~~~~~~~~~ 10~~~~~~~~~0.0 \\

$^o{B}$~~~~~~~~~ 9 ~~~~~~~~~ 13.1 & $^o{B}$~~~~~~~~~ 9~~~~~~~~~ 3.4
& $^o{A}$~~~~~~~~~ 9~~~~~~~~~ 39.6 & $^o{A}$~~~~~~~~~ 9~~~~~~~~~ 42.4 \\

$^o{A}$~~~~~~~~~ 9 ~~~~~~~~~ 26.1 & $^e{A}$~~~~~~~~~ 8~~~~~~~~~ 10.2
& $^o{B}$~~~~~~~~~ 9 ~~~~~~~~~ 54.2 & $^o{B}$~~~~~~~~~ 9~~~~~~~~~ 58.8 \\

$^e{A}$~~~~~~~~~ 8 ~~~~~~~~~ 27.3 & $^o{B}$~~~~~~~~~ 7~~~~~~~~~ 20.1
& $^o{B}$~~~~~~~~~ 9 ~~~~~~~~~ 62.4 & $^o{B}$~~~~~~~~~ 9~~~~~~~~~ 69.4 \\
\hline 
\end{tabular}
\end{center}

\newpage

\noindent{\bf Figure Captions}
\vspace{0.5cm}

\noindent {1.} Representative $\rm M_s=0$ state in (a) 6 spin-$1/2$ cluster, 
(b) $\rm Mn_{12}Ac$ cluster with first four sites each having spin $\rm S=3/2$ 
and the remaining eight sites each having spin $\rm S=2$. Numbers in 
parenthesis correspond to the $\rm M_s=0$ value at the site. The bit 
representations as well as the integer values are given just below the 
diagrams.

\noindent {2.} A schematic diagram of the exchange interactions between the Mn 
ions in the $\rm Mn_{12}Ac$ molecule.

\noindent {3.} Spin density of $\rm Mn_{12}Ac$ for parameter values: $J_1$=
215K, $J_2$=85K, $J_3$=85K and $J_4$=-64.5K. (a) Spin density for ground state 
(S=10,$\rm M_s$=10). (b) Spin density for 1$^{st}$ excited state (S=9, $\rm 
M_s$=9).

\noindent {4.} A schematic diagram of the exchange interactions between the Fe 
ions in the $\rm Fe_8$ molecule.

\noindent {5.} Spin density of $\rm Fe_8$ for parameter values: J$_1$ = 150K,
J$_2$ = 25K, J$_3$ = 30K, J$_4$ = 50K. (a) Spin density for ground state 
(S=10,$\rm M_s$=10). (b) Spin density for 1$^{st}$ excited state (S=9, $\rm 
M_s$=9). \\ 
\noindent 
Spin density of $\rm Fe_8$ for J$_1$ = 180K, J$_2$ = 153K, J$_3$ = 22.5K, 
J$_4$ = 52.5K parameter values. (c) Spin density for ground state (S=10,$\rm 
M_s$=-10). (d) Spin density for 1$^{st}$ excited state (S=9,$\rm M_s$=9). \\
\noindent 
Spin density of $\rm Fe_8$ for J$_1$ = 195K, J$_2$ = 30K, J$_3$ = 52.5K, 
J$_4$ = 22.5K parameter values. (e) Spin density for ground state (S=10,
$\rm M_s$=10). (f) Spin density for 1$^{st}$ excited state (S=9,$\rm M_s$ =9).
\noindent 
Spin density of $\rm Fe_8$ for J$_1$ = 201K, J$_2$ = 36.2K, J$_3$ = 58.3K, 
J$_4$ = 26.1K parameter values. (g) Spin density for ground state (S=10,
$\rm M_s$=10). (h) Spin density for 1$^{st}$ excited state (S=9,$\rm M_s$ =9).

\noindent {6.} A schematic diagram of the exchange interactions between the V 
ions in the $\rm V_{15}$ molecule.

\noindent {7.} Spin density of $\rm V_{15}$ for J$_1$ = 800K, J$_2$ = 300K, 
J$_3$ = 150K parameter values. (a) Spin density in one of the ground states 
(S=0.5,$\rm M_s$=0.5). (b) Spin density for excited state (S=1.5, $\rm M_s$=
1.5).

\begin{center}
\begin{figure}
\vspace*{-4cm}
\hspace*{-4cm}
\epsfig{file=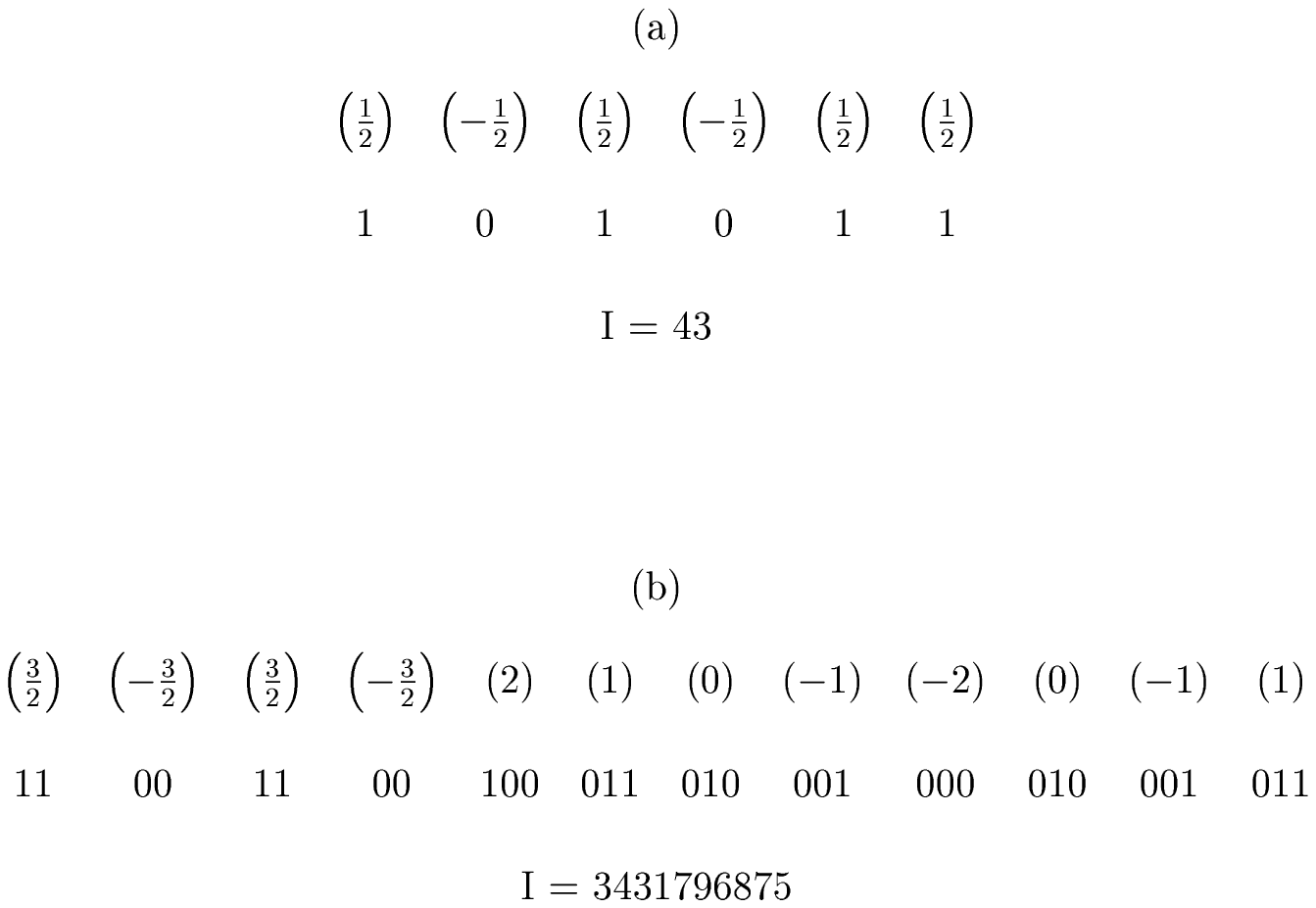}
\label{fig1}
\vspace*{-10cm}
\centerline{\LARGE {Fig. 1}} 
\end{figure}
\end{center}

\begin{center}
\begin{figure}
\epsfig{file=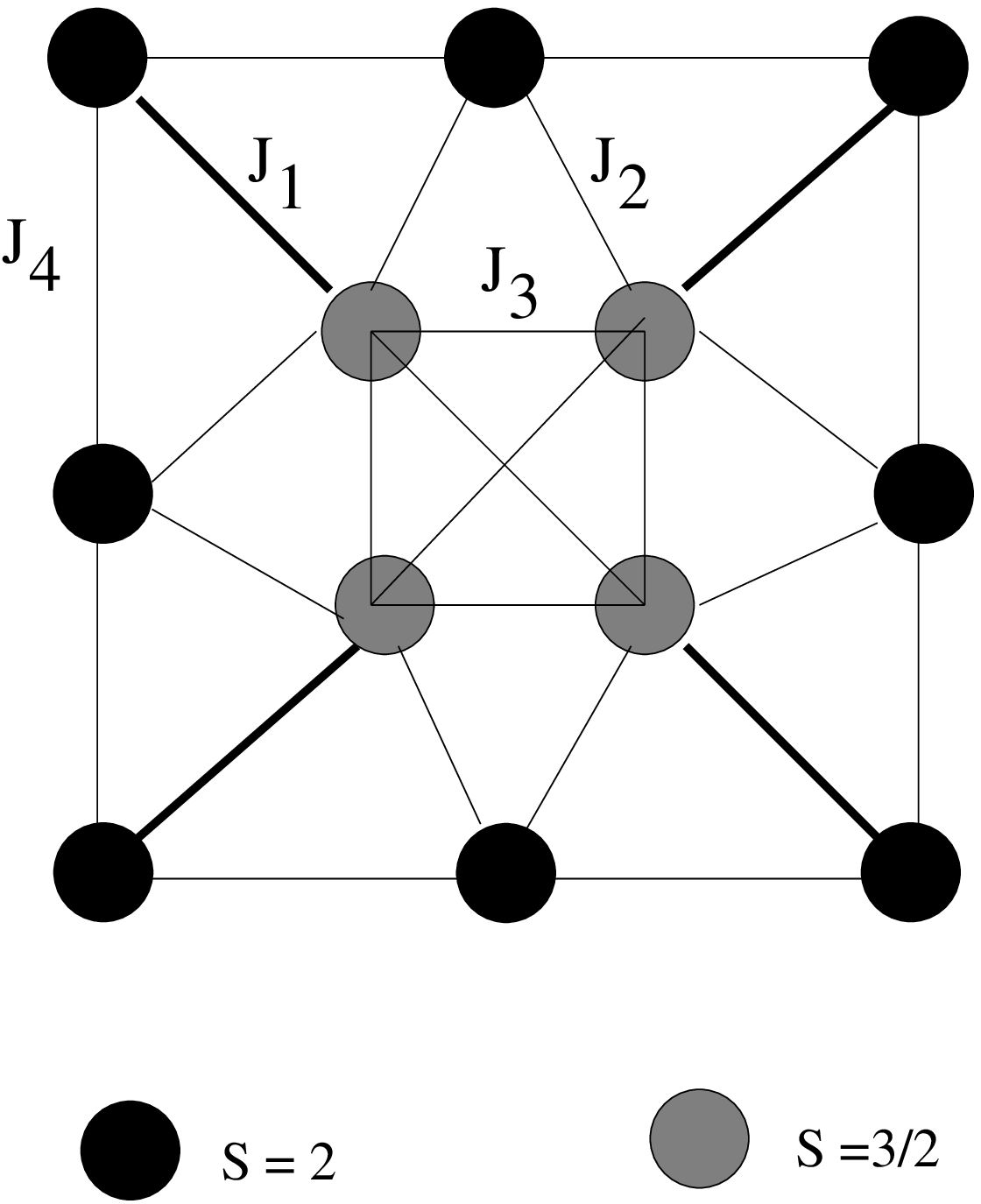}
\label{fig2}
\vspace{2cm}
\centerline{\LARGE {Fig. 2}} 
\end{figure}
\end{center}

\begin{center}
\begin{figure}
\vspace{3cm}
\epsfig{file=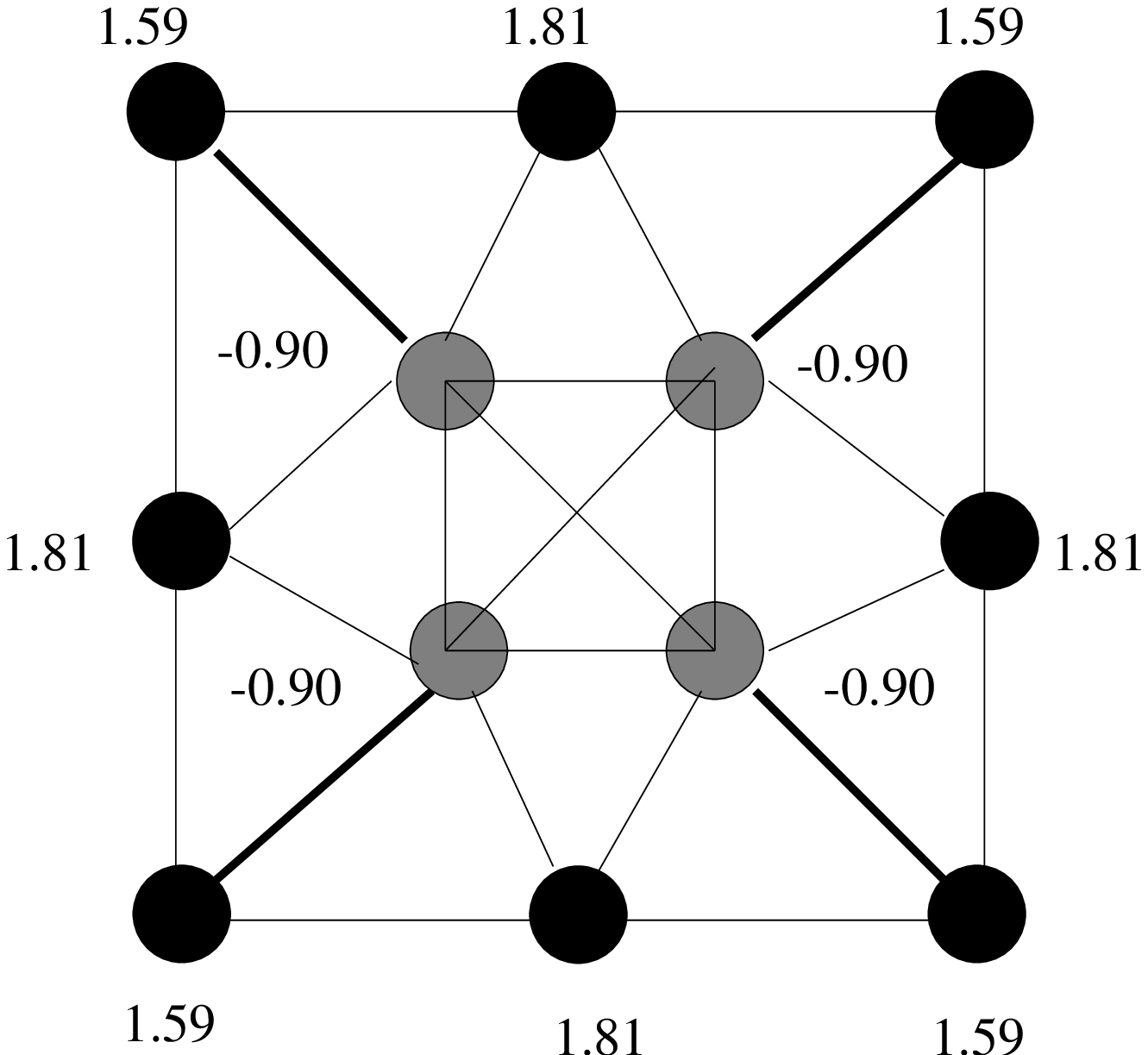}
\label{fig3a}
\vspace{2cm}
\centerline{\LARGE {Fig. 3(a)}} 
\end{figure}
\end{center}

\begin{center}
\begin{figure}
\vspace{3cm}
\epsfig{file=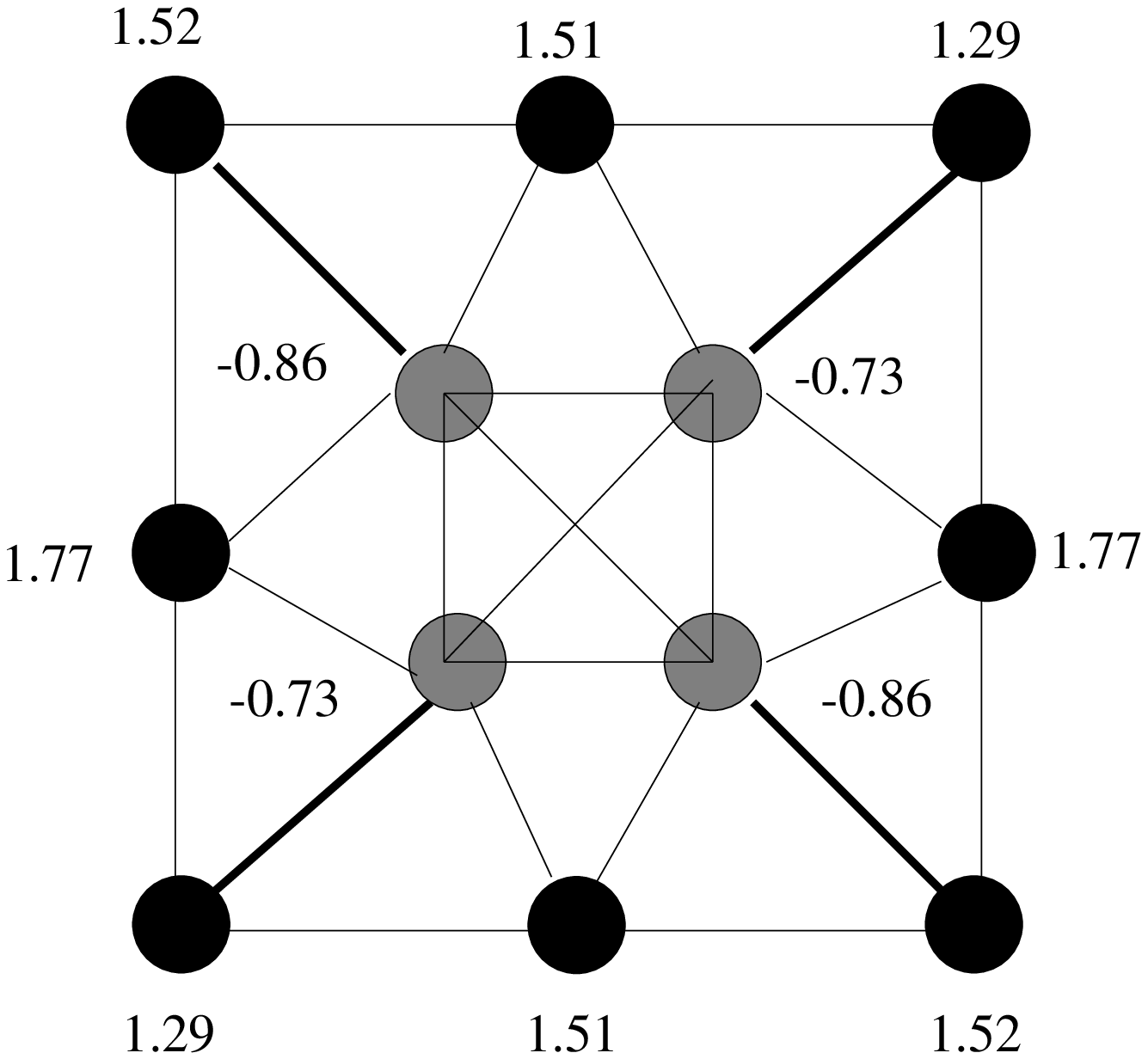}
\label{fig3b}
\vspace{2cm}
\centerline{\LARGE {Fig. 3(b)}}
\end{figure}
\end{center}

\begin{center}
\begin{figure}
\vspace{3cm}
\epsfig{file=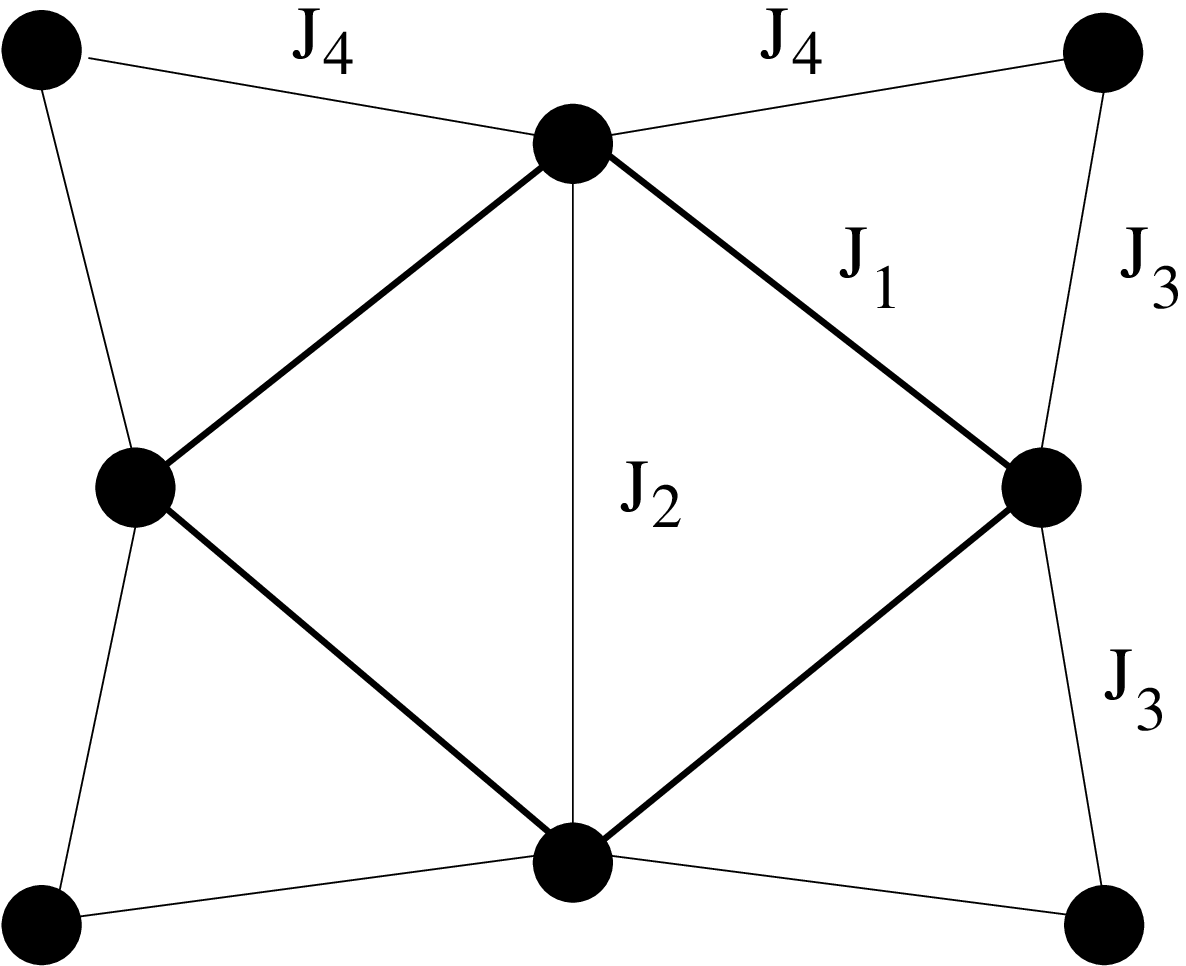}
\label{fig4}
\vspace{2cm}
\centerline{\LARGE {Fig. 4}} 
\end{figure}
\end{center}

\begin{center}
\begin{figure}
\vspace{3cm}
\epsfig{file=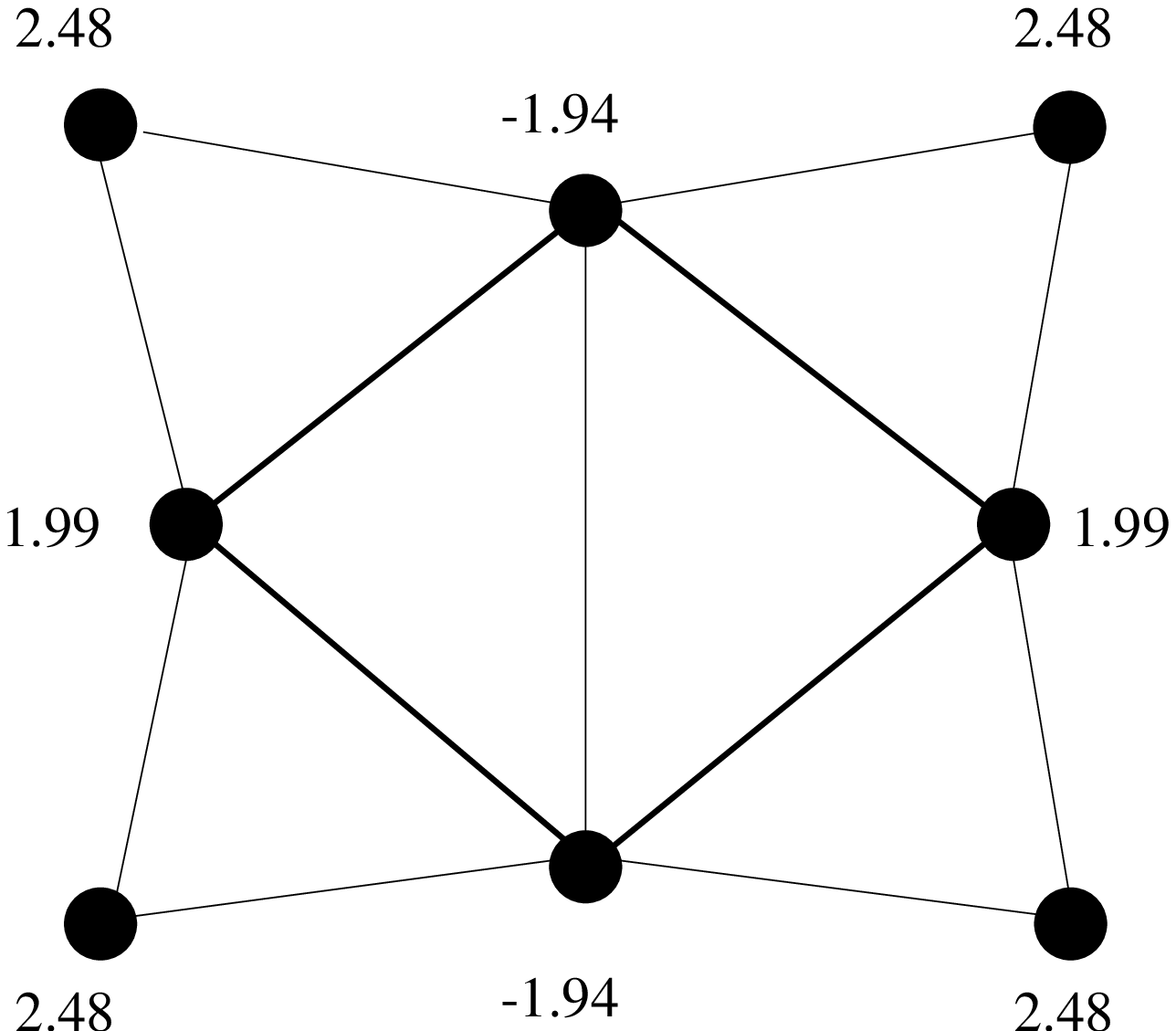}
\label{fig5a}
\vspace{2cm}
\centerline{\LARGE {Fig. 5(a)}} 
\end{figure}
\end{center}

\begin{center}
\begin{figure}
\vspace{3cm}
\epsfig{file=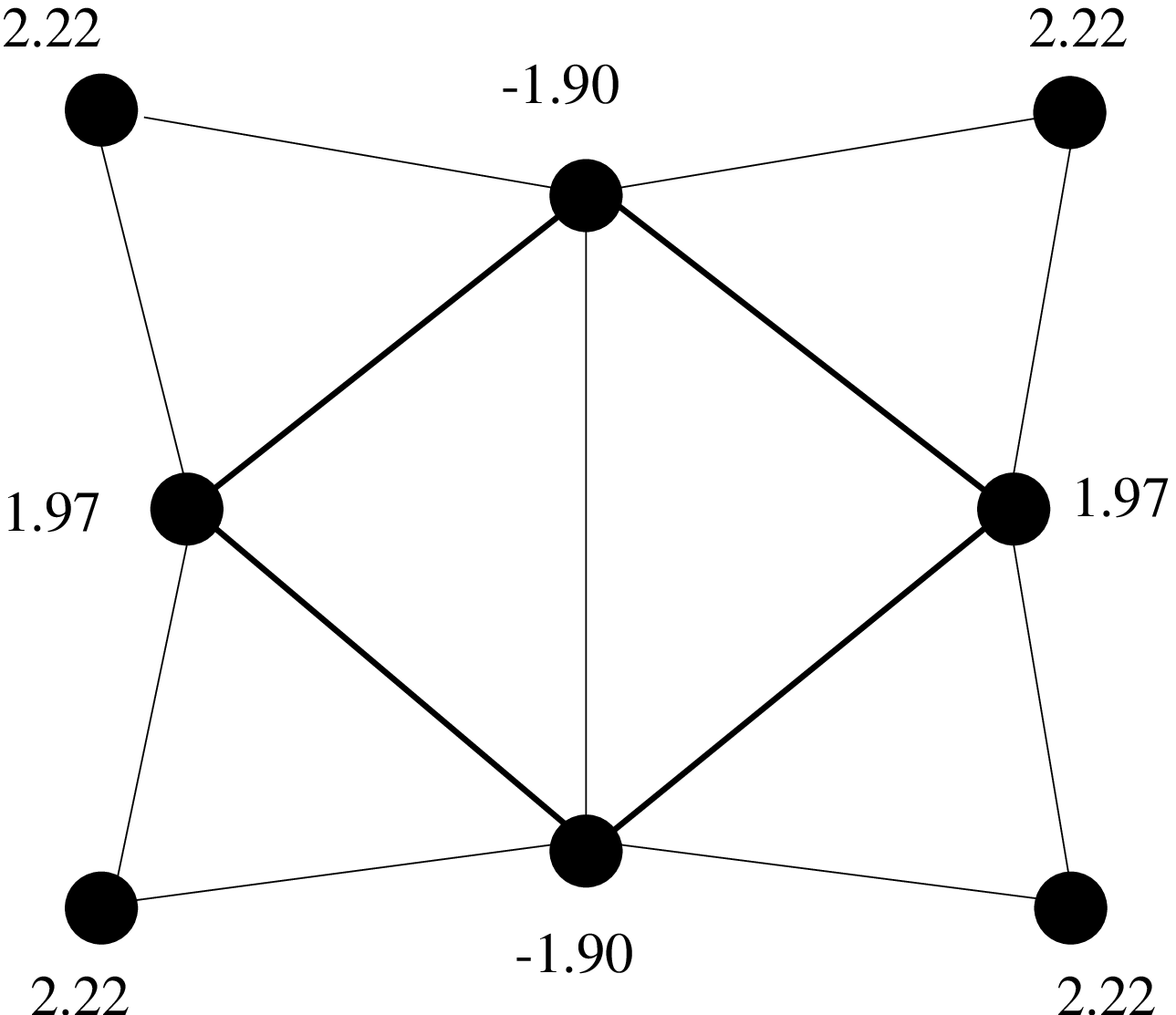}
\label{fig5b}
\vspace{2cm}
\centerline{\LARGE {Fig. 5(b)}} 
\end{figure}
\end{center}

\begin{center}
\begin{figure}
\vspace{3cm}
\epsfig{file=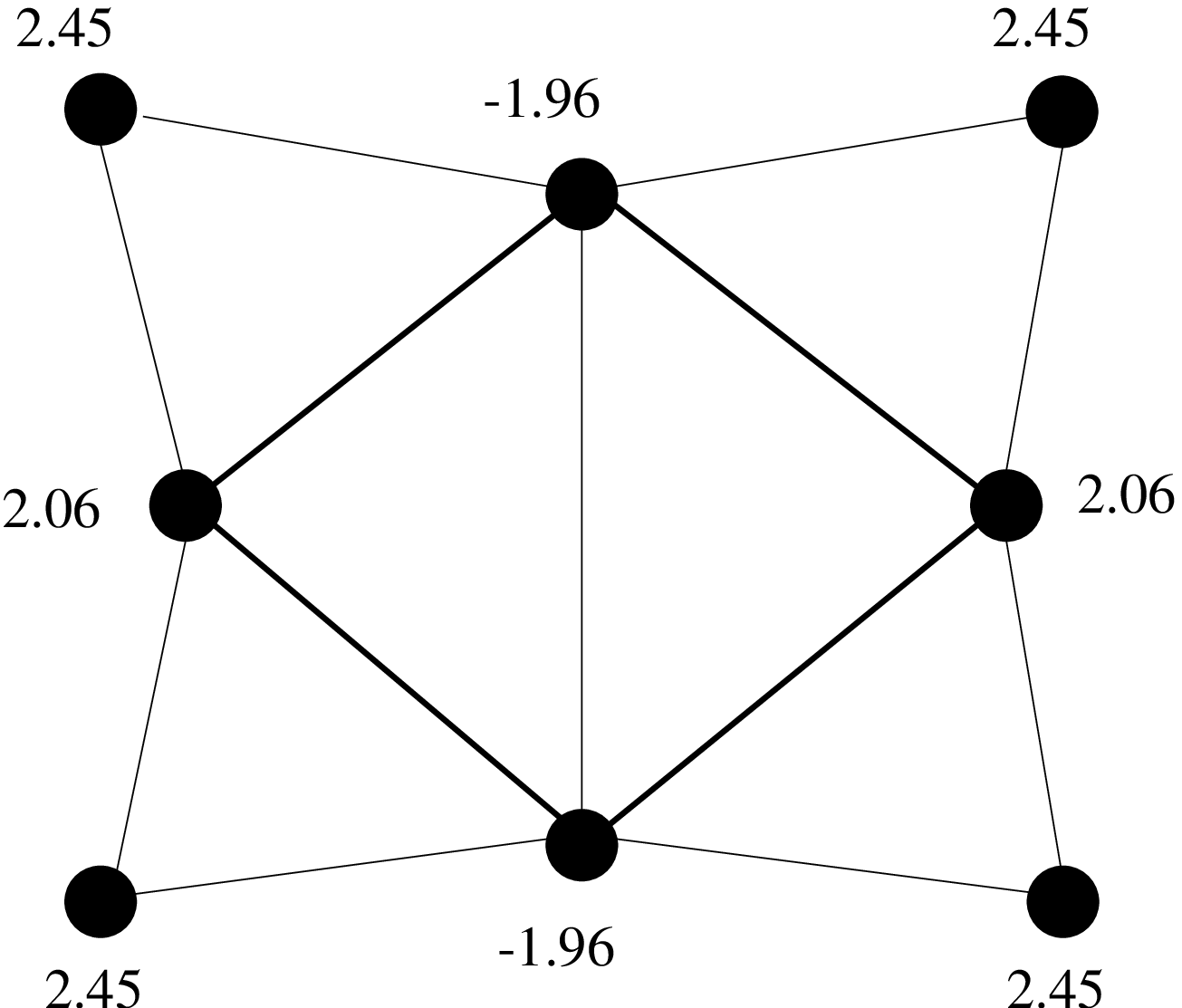}
\label{fig5c}
\vspace{2cm}
\centerline{\LARGE {Fig. 5(c)}} 
\end{figure}
\end{center}

\begin{center}
\begin{figure}
\vspace{3cm}
\epsfig{file=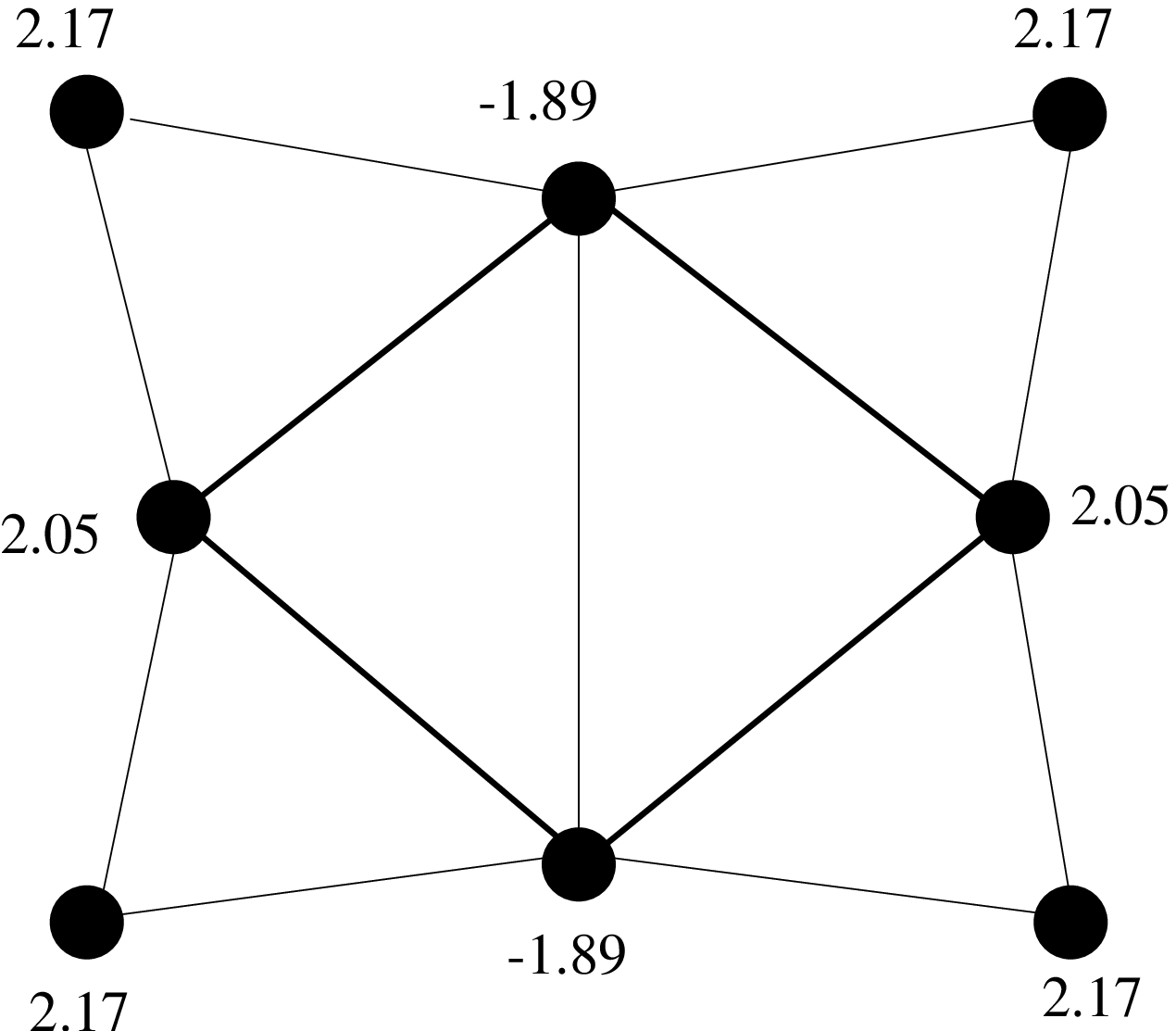}
\label{fig5d}
\vspace{2cm}
\centerline{\LARGE {Fig. 5(d)}} 
\end{figure}
\end{center}

\begin{center}
\begin{figure}
\vspace{3cm}
\epsfig{file=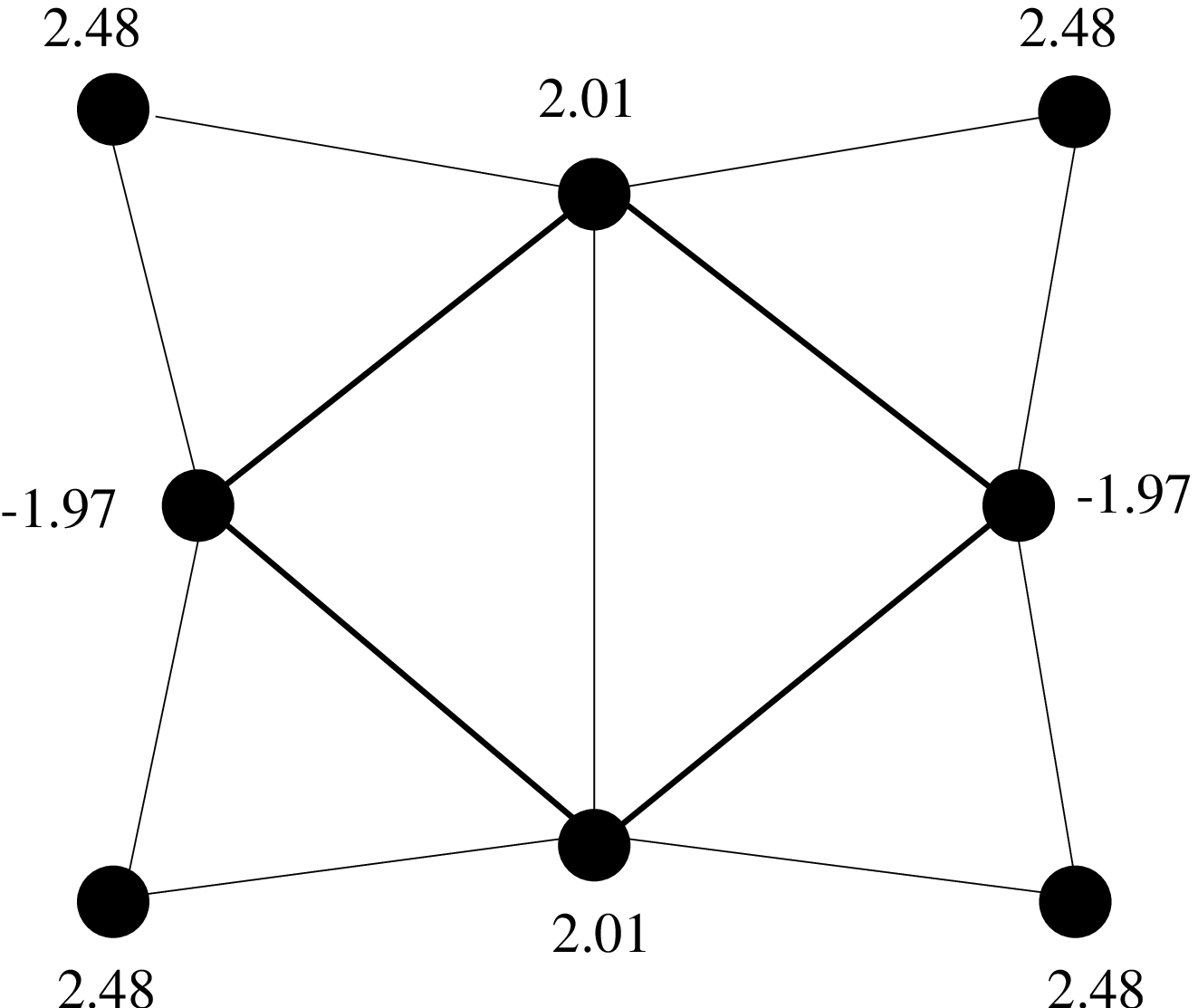}
\label{fig5e}
\vspace{2cm}
\centerline{\LARGE {Fig. 5(e)}} 
\end{figure}
\end{center}

\begin{center}
\begin{figure}
\vspace{3cm}
\epsfig{file=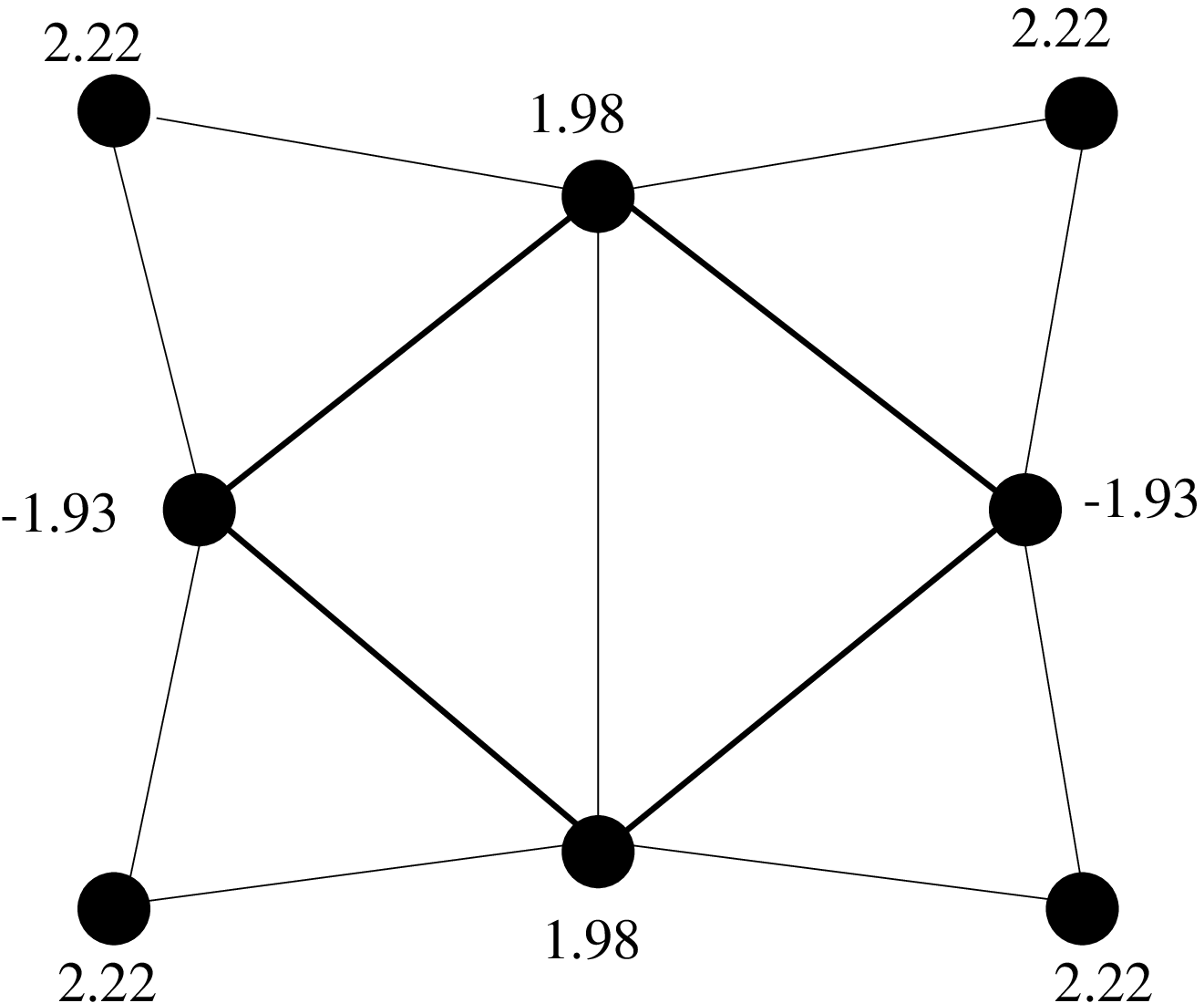}
\label{fig5f}
\vspace{2cm}
\centerline{\LARGE {Fig. 5(f)}} 
\end{figure}
\end{center}

\begin{center}
\begin{figure}
\vspace{3cm}
\epsfig{file=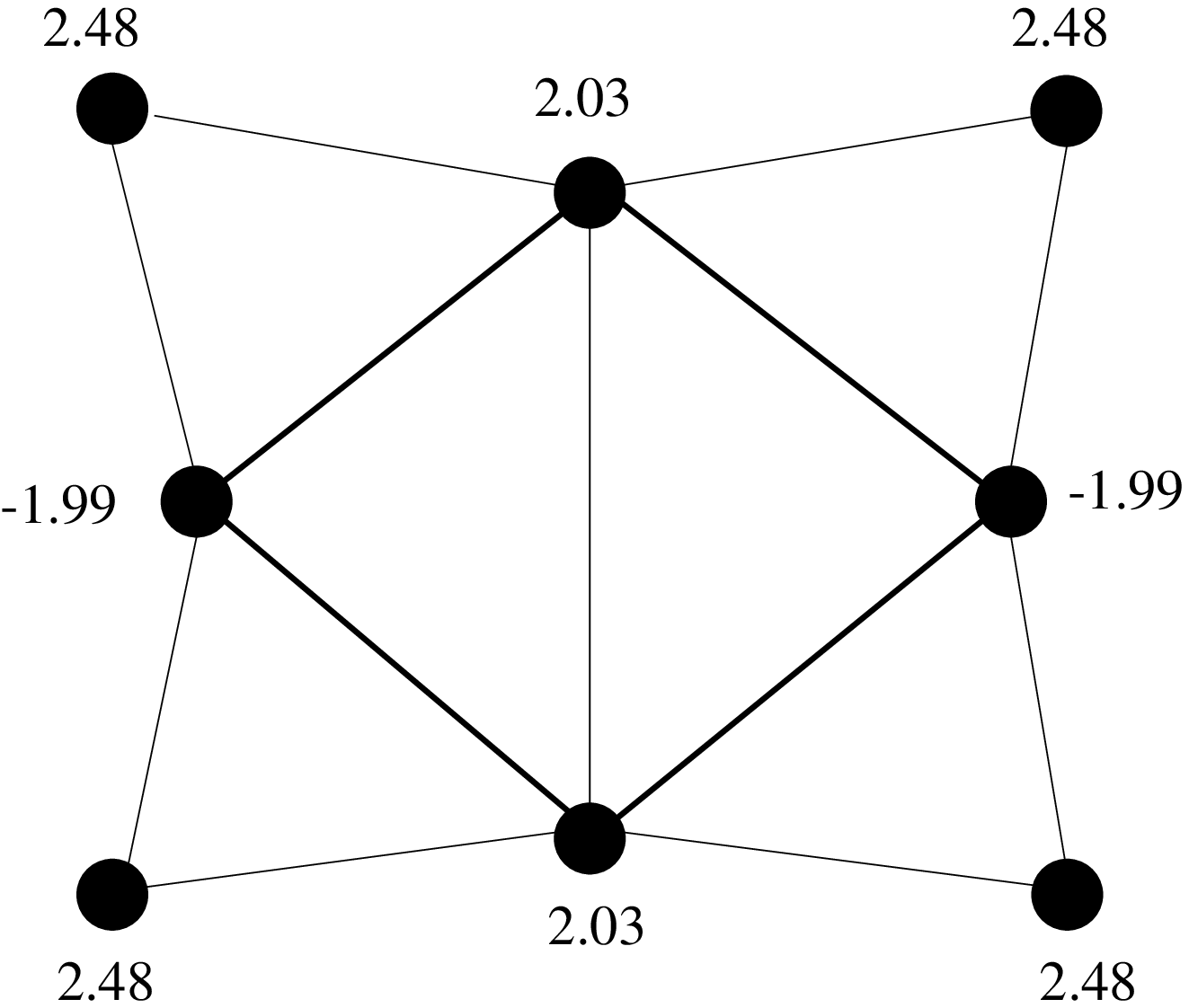}
\label{fig5g}
\vspace{2cm}
\centerline{\LARGE {Fig. 5(g)}} 
\end{figure}
\end{center}

\begin{center}
\begin{figure}
\vspace{3cm}
\epsfig{file=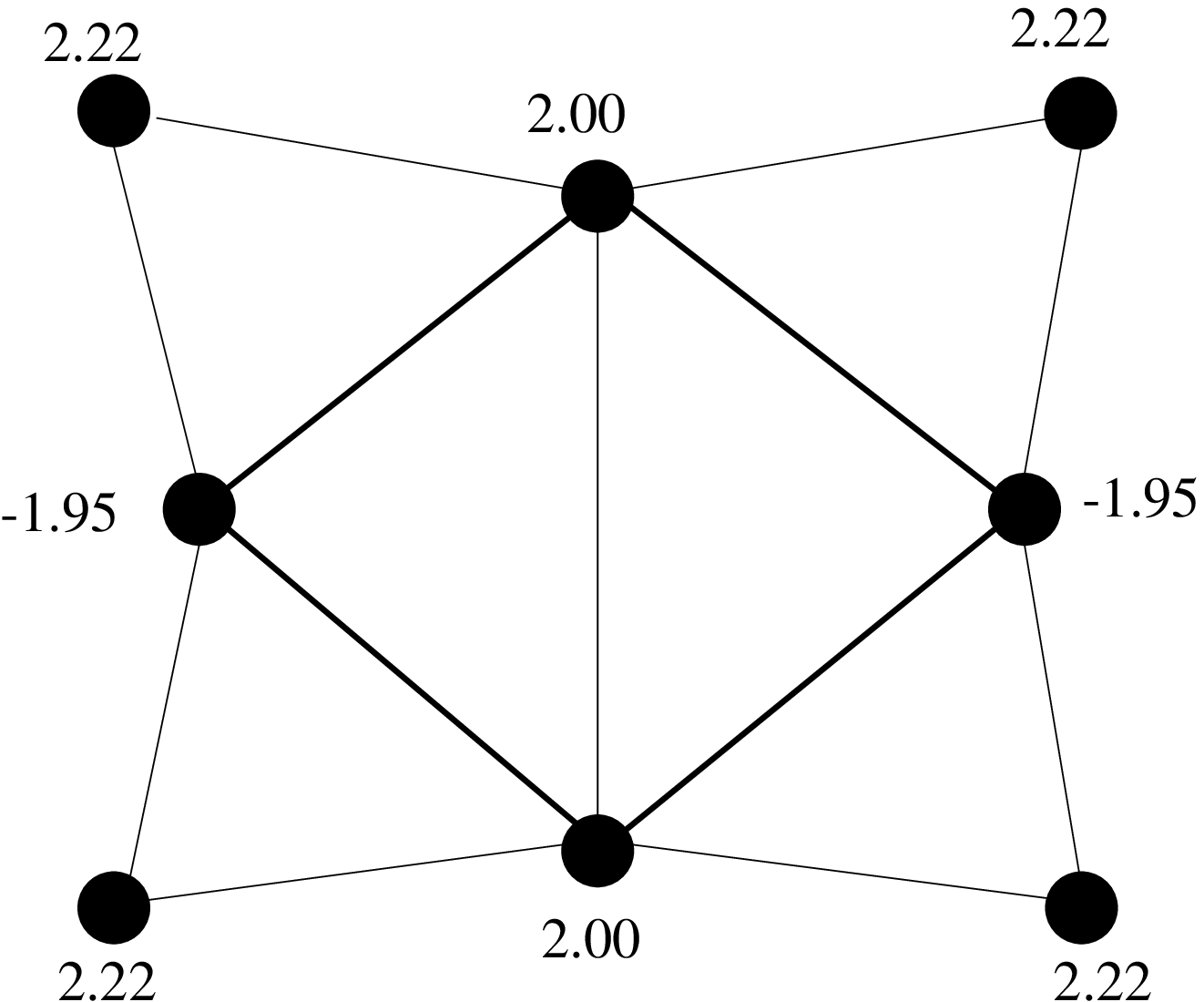}
\label{fig5h}
\vspace{2cm}
\centerline{\LARGE {Fig. 5(h)}} 
\end{figure}
\end{center}

\begin{center}
\begin{figure}
\epsfig{file=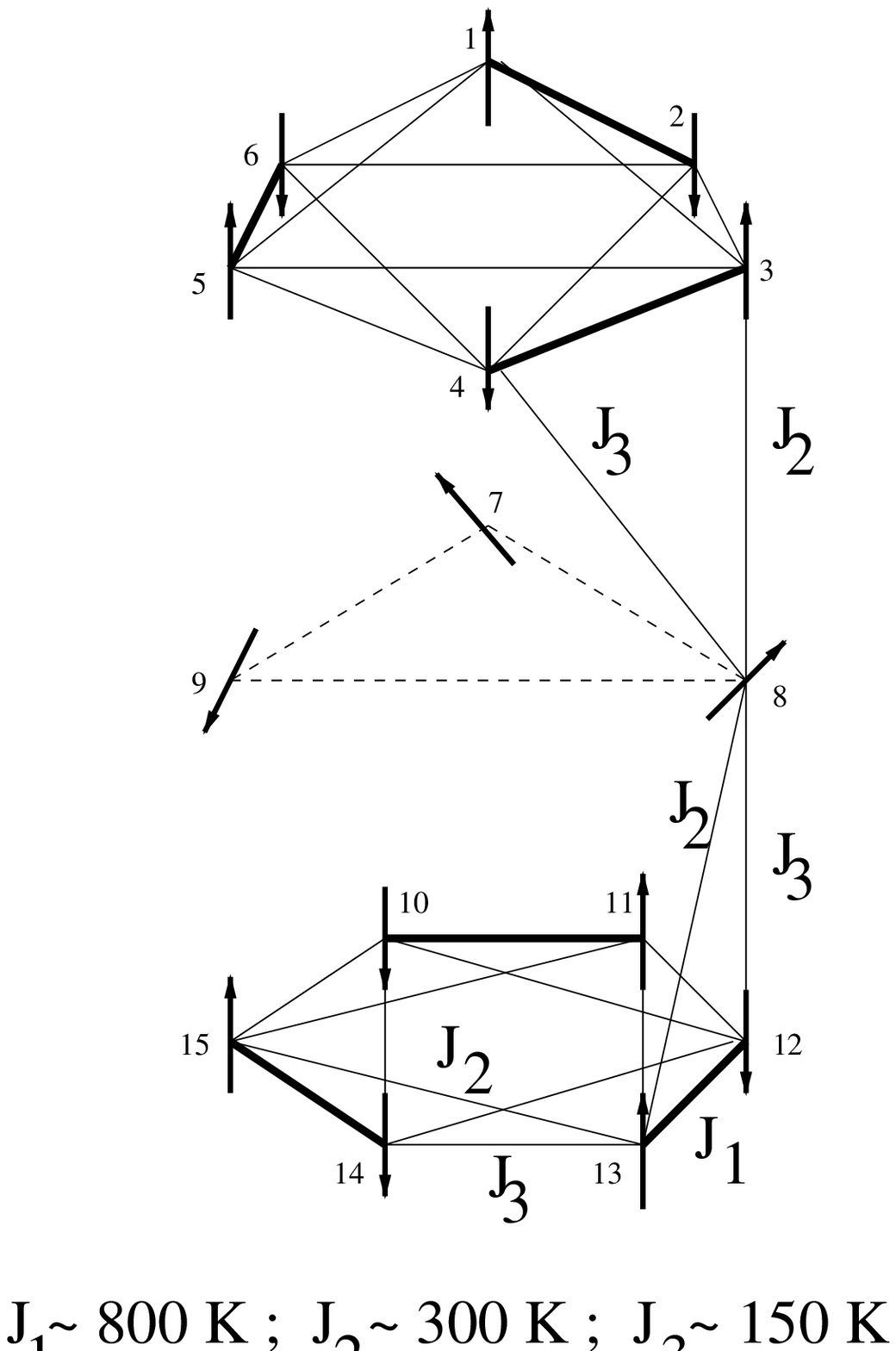}
\label{fig6}
\vspace*{1cm}
\centerline{\LARGE {Fig. 6}} 
\end{figure}
\end{center}

\begin{center}
\begin{figure}
\epsfig{file=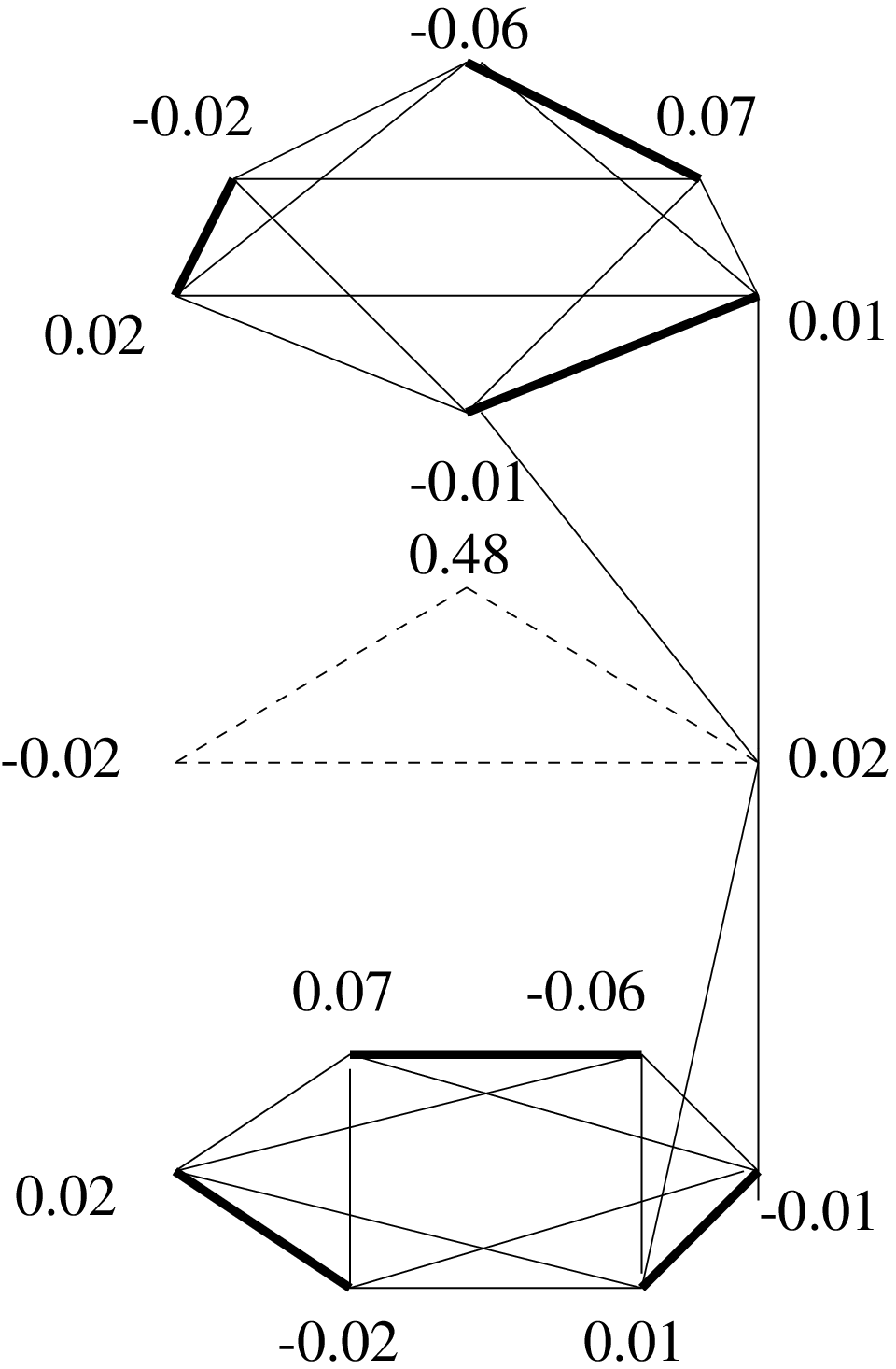}
\label{fig7a}
\vspace{2cm}
\centerline{\LARGE {Fig. 7(a)}} 
\end{figure}
\end{center}

\begin{center}
\begin{figure}
\epsfig{file=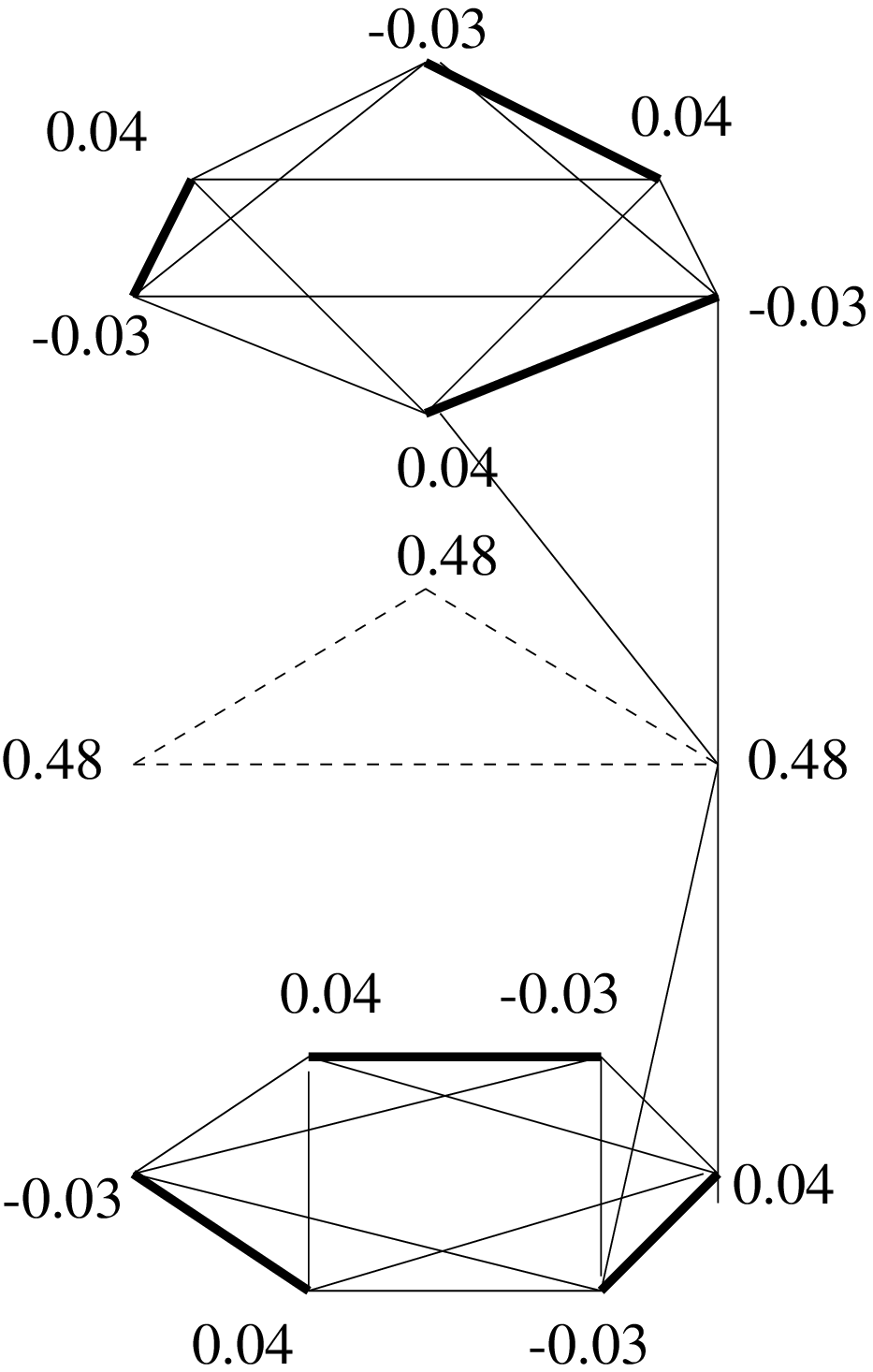}
\label{fig7b}
\vspace{2cm}
\centerline{\LARGE {Fig. 7(b)}} 
\end{figure}
\end{center}

\end{document}